\documentclass[twocolumn]{aastex701}

\usepackage{amsmath}
\usepackage[utf8]{inputenc}
\usepackage{graphicx}

\usepackage{lipsum}

\newcommand{\dif}{\mathrm{d}}
\newcommand{\e}{\mathrm{e}}

\newcommand{\Myr}{\,\mathrm{Myr}}
\newcommand{\Gyr}{\,\mathrm{Gyr}}
\newcommand{\AU}{\,\mathrm{AU}}
\newcommand{\pc}{\,\mathrm{pc}}

\newcommand{\MSol}{M_{\odot}}
\newcommand{\RSol}{R_{\odot}}

\newcommand{\cmc}{{\tt CMC}}

\newcommand{\ciera}{\affiliation{Center for Interdisciplinary Exploration and Research in Astrophysics (CIERA), Northwestern University, 1800 Sherman Ave., Evanston, IL 60201, USA}}
\newcommand{\nupa}{\affiliation{Department of Physics and Astronomy, Northwestern University, 2145 Sheridan Rd., Evanston, IL 60208, USA}}
\newcommand{\ucsd}{\affiliation{Department of Astronomy and Astrophysics, University of California, San Diego, La Jolla, CA 92093, USA}}

\shorttitle{Binaries in globular clusters}
\shortauthors{O'Connor, Kremer \& Rasio}

\allowdisplaybreaks

\defcitealias{Kremer2020catalog}{K20}
\defcitealias{Offner2023binaries}{O23}

\received{2026 March 30}
\submitjournal{ApJ}

\begin{document}

\title{An analytical approach to binary populations in globular clusters}

\author[0000-0003-3987-3776, sname='O\'Connor', gname='Christopher E.']{Christopher E.\ O'Connor}
\ciera
\email[show]{christopher.oconnor@northwestern.edu}

\author[0000-0002-4086-3180, sname='Kremer', gname='Kyle']{Kyle Kremer}
\ucsd
\email{kykremer@ucsd.edu}

\author[0000-0002-7132-418X,gname='Frederic A.',sname='Rasio']{Frederic A.\ Rasio}
\ciera
\nupa
\email{rasio@northwestern.edu}

\correspondingauthor{Chris O'Connor}

\begin{abstract}
    Globular clusters (GCs) display much lower binary fractions than found among main-sequence stars in the solar neighborhood. 
    The physical cause of this difference is debatable: 
    does it reflect different star formation outcomes at low metallicity and/or high density, 
    the dynamical processing of primordial binaries over cluster lifetimes, or a combination of the two? 
    Starting from the assumption that the initial binary distribution (IBD) in GCs is the same as the binary distribution observed in the solar neighborhood, 
    we show with straightforward analytical calculations that the dynamical dissolution of ``soft'' primordial binaries can fully explain the main-sequence binary fractions in present-day GCs. 
    We validate our estimates against a detailed $N$-body simulation with the {\tt Cluster Monte Carlo} code. 
    Adopting the view that the observed binary fraction in a given cluster constrains the location of the hard/soft boundary at birth, 
    we infer that surviving Milky Way GCs had a similar distribution of birth radii to young massive clusters in the local universe. 
    Our findings underscore the crucial role of stellar black holes (through ``black hole burning'') in sculpting GC binary populations 
    and reinforce the need for realistic initial conditions in theoretical modeling of GC dynamics.
\end{abstract}


\section{Introduction} \label{s:Intro}

Stellar multiplicity is the norm across most galactic environments. 
A major exception to this rule occurs among low-mass main-sequence stars and giants in old stellar systems such as globular clusters (GCs).
Observational studies spanning several decades, and employing various techniques, have consistently found 
much lower binary fractions within well-studied GCs than in the Galactic field and halo, 
typically $\sim 1 \mbox{--} 10\%$ in the former \citep[e.g.,][]{Albrow2001_47tucbinaries, Sollima+2007, Milone2012gcbinaryfraction, Ji2015_gcbinaries, MullerHorn2025_47tucbinaries} versus $\sim 50 \%$ in the latter \citep[e.g.,][]{DM1991, Raghavan+2010, Moe2019metallicity, Offner2023binaries}. 
A similar paucity of low-mass binaries may exist in GC stellar streams \citep{UsmanJi2024gcstreams, Phillips+2026}. 
At face value, the intuitive explanation for this trend is that dynamical interactions within GCs reduce the total binary fraction from its primordial value \citep[e.g.,][]{Hut+1992hardbinaries, Ivanova+2005, Fregeau+2009, Leigh+2013, Leigh+2015, Wang+2022, Phillips+2026}. 
On the other hand, many early studies were interested in how to \emph{form} long-lived binaries in GCs, 
especially to explain the presence of low-mass X-ray binaries and millisecond pulsars in these environments \citep[e.g.,][]{Fabian+1975, McMillan+1987, Hut+1992hardbinaries}.
Clearly, demonstrating end-to-end consistency for any version of binary evolution in GCs is a tall order. 
Several aspects of the problem call for re-examination in light of recent developments.  

The gold standard for testing theories of long-term cluster evolution is set by comparison between direct-summation $N$-body simulations and observations of Milky Way GCs \citep[e.g.,][]{BaumgardtHilker2018}. 
However, full-lifetime direct integrations of dense clusters with large primordial binary populations 
have only recently become computationally feasible (e.g., \citealt{Wang2020petar, Wang+2022, Kamlah+2022}; 
see \citealt{SpurzemKamlah2023} for a recent review of direct $N$-body methods), 
and direct integrations of clusters with $\gtrsim 10^{6}$ stars (such as typical GCs) remain costly and rare \citep{Wu+2025}. 
Monte Carlo methods can be deployed to study in-cluster binary evolution in higher-mass GCs 
\citep[e.g.,][]{Fregeau+2009, Leigh+2015, Hong+2015, Hong+2017, Hong+2018, Gonzalez2021, GonzalezPrieto2022, GonzalezPrieto2024, Kiroglu2025bhaccretion}. 
Even so, observations of stellar binaries in the nearest analogs of GC progenitors -- young massive clusters (YMCs) and super star clusters (SSCs) -- 
are sensitive only to the most massive systems \citep[e.g.,][]{Sana+2012, Sana2013_30Dor, Sana2025bloem, Clark+2020, Clark2023arches, Ritchie2022w1obbinaries}. 
Options for direct observational studies of intermediate- and low-mass binaries in high-density environments are scant.

Additionally, the demographics of primordial binaries as a function of stellar mass is not known {\it a priori.} 
Like the stellar initial mass function (IMF), the initial binary distribution (IBD) might depend in a complicated manner 
on the physics of star formation in an environment of a given density and metallicity \citep[e.g.,][]{Grudic+2021, Guszejnov+2022, Guszejnov+2023, Matzner2024}.  
Binary interactions during the pre-MS phase may likewise complicate definitions of ``primordial''/``initial'' binary properties \citep[e.g.,][]{Kroupa1995}. 
Only recently has there been sufficient data on binary demographics at all separations 
to provide a realistic empirical template for the IBD of the Galactic field population \citep[e.g.,][]{MoeDiStefano2017, Moe2019metallicity, ElBadry2019widebinmetallicity, Hwang+2021, Offner2023binaries}. 

Faced with these uncertainties, there is value in the old-fashioned exercise of postulating 
a universal IBD and then assessing the extent to which observations require variations between different environments \citep[e.g.,][]{Leigh+2015, Belloni+2017}. 
This channels the spirit in which earlier works have interrogated the hypothesis of a universal stellar IMF \citep[e.g.,][]{Kroupa2001imf, Chabrier2003} 
and, in some instances, inferred variations of the IMF in certain environments \citep[e.g.,][]{Bartko+2010, Lu+2013, Jerabkova2017, Hosek2019, vanDokkumConroy2024}.

In this study, we advance the case for a universal IBD between the solar neighborhood and Galactic GCs. 
In Section \ref{s:ObsSummary}, we summarize pertinent observational constraints on field and cluster binary populations from the literature. 
In Section \ref{s:Analytics}, we present a concise analytical formalism to account for the effects of cluster dynamics on primordial binaries;
we predict the correct range of observed binary fractions among low-mass stars in well-studied GCs 
by invoking the dissolution of dynamically ``soft'' binaries -- 
those with orbital velocities below the cluster's velocity dispersion at birth. 
In Section \ref{s:Dynamics}, we quantify the rate and energy cost of soft binary dissolution with simple analytical estimates, 
confirm the overall picture with a detailed $N$-body simulation, and infer birth conditions for a set of Milky Way GCs based on our results. 
In Section \ref{s:Discuss}, we discuss how our results can inform future theoretical and observational investigations of binaries in dense clusters, 
as well as how they relate to ideas developed in previous works. 
In Section \ref{s:Summary}, we summarize our takeaway messages.

\section{Pertinent Observations} \label{s:ObsSummary}

Here we give a brief synopsis of current knowledge about observed binary populations in GCs and the field of the solar neighborhood, 
and use these data to formulate a fiducial universal IBD for the Milky Way.

Early searches for stellar binaries in GCs using traditional spectroscopic or photometric monitoring
were rather limited in the scope, 
but nonetheless established that binaries are much rarer in GCs than in the field \citep[e.g.,][]{Albrow2001_47tucbinaries}. 
In recent years, the most widely employed method of searching for binaries in a GC is to measure the broadening of the cluster's main sequence on a color--magnitude diagram \citep[e.g.,][]{Sollima+2007, Milone2012gcbinaryfraction, Ji2015_gcbinaries, MullerHorn2025_47tucbinaries}. 
This technique is sensitive to the frequency of unresolved, near-equal-mass binaries (mass ratios $q \gtrsim 0.5$), 
but is subject to contamination in crowded fields near the center of a cluster
and to modeling uncertainties related to how one corrects for the low-$q$ binaries to which it is not sensitive. 
Bearing these limitations in mind, the photometric-broadening method has yielded consistent estimates for well-studied GCs over the past few decades, 
with total binary fractions of $\sim 1 \mbox{--} 10\%$ estimated near the half-mass or half-light radius 
\citep[e.g.,][]{Milone2012gcbinaryfraction, Ji2015_gcbinaries, MullerHorn2025_47tucbinaries}. 
A GC's binary fraction tends to increase towards its center, typically by a factor of $\sim 2 \mbox{--} 5$ above the half-mass value \citep[e.g.,][]{Sollima+2007, Milone2012gcbinaryfraction, Ji2015_gcbinaries}. 
This is the expected result of mass segregation via two-body relaxation \citep{Hut+1992hardbinaries, HeggieHut2003}. 
The minority of GCs for which larger total binary fractions ($\sim 15 \mbox{--} 30 \%$) are measured 
stand out as unusually diffuse, which will prove consistent with the calculations of Section \ref{s:Analytics}. 

A few other lines of evidence have opened up in recent years, thanks to {\it Gaia}. 
Spectroscopic observations with {\it Gaia}'s Radial Velocity Spectrometer 
have yielded independent estimates of binary fractions for a subset of nearby GCs, which largely agree with the photometrically derived values \citep{BashBelokurov2025}. 
Meanwhile, {\it Gaia} astrometry has facilitated the identification of GC stellar streams, 
whose binary fractions also appear to be low \citep{UsmanJi2024gcstreams}. 

The binary population in the solar neighborhood, including nearby resolved star-forming regions, is much better understood. 
The fraction of primary stars with at least one bound stellar companion, or ``multiplicity fraction'' (MF) 
rises monotonically from $\sim 20\%$ at the lowest masses to virtually $100\%$ at the highest (see \citealt{Offner2023binaries} and references therein). 
We adopt the bias-corrected MF as compiled in Table 1 of \citet{Offner2023binaries} 
for primary stellar masses of $0.075 \mbox{--} 50 \MSol$; 
these data are described well by the following sigmoid-like fitting formula:
\begin{equation} \label{eq:offner_mf_fit}
    F_{\rm b,tot}(m_{1}) = F_{\rm lo} + \frac{F_{\rm hi} - F_{\rm lo}}{1 + (m_{1}/m_{\rm mid})^{-\beta}},
\end{equation}
with best-fitting parameters $F_{\rm lo} = 0.18$, $F_{\rm hi} = 0.99$, $m_{\rm mid} = 1.6 \MSol$, and $\beta = 3.1$. 
We use this function as an Ansatz for the total primordial binary fraction as a function of stellar-mass. 

The second step in building the IBD is to decide the separation (i.e., period) distribution of primordial binaries. 
This determines the predicted proportions of hard and soft binaries in a given cluster. 
The observed separation distribution of solar-neighborhood binaries varies with mass 
(and also with the occurrence of triple and higher-order multiple systems, which we do not take into account). 
Low-mass binaries follow a roughly log-normal separation distribution 
peaked at $\sim 10 \mbox{--} 30 \AU$ and with a dispersion of $\sim 1 \mbox{--} 1.5 \, {\rm dex}$ \citep[e.g.,][]{DM1991, Raghavan+2010}. 
Companions to high-mass stars are closer to being flat in the log of separation overall \citep[e.g.,][]{Sana+2012, Offner2023binaries}, 
though there is also an overabundance of close companions due to the prevalence of triple systems in observed massive-star samples. 
We consider both log-flat and log-normal separation distributions in our calculations in Section \ref{s:Analytics}, 
finding that the result depends moderately on the form of the distribution for high-mass systems but only weakly for low-mass.

Observational input on the mass-ratio and eccentricity distributions in multi-star systems 
is also required to fully realize a model binary population. 
Both distributions have been characterized in some detail as functions of primary mass and separation \citep[e.g.,][]{MoeDiStefano2017}. 
In a dense cluster, strong dynamical interactions efficiently thermalize binary eccentricities, 
and exchange interactions skew the mass-ratio distribution toward near-equal-mass systems \citep[e.g.,][]{Heggie1975, Hut+1992hardbinaries}.  
Only in a minority of very close and/or very massive binaries, 
where tidal interactions operate on a timescale that competes with the rate of strong dynamical encounters, 
might the initial mass-ratio or eccentricity distribution make an observationally relevant impact on the population. 
For purposes of this work, it is sufficient to consider only the total binary fraction and separation distribution as functions of primary stellar mass. 

\section{Analytical formalism} \label{s:Analytics}

\subsection{Preliminaries} \label{s:Analytics:Prelim}

Consider a young star cluster immediately after the expulsion of its residual gas, modeled as a Plummer sphere of mass $M_{\rm cl}$ and scale length $b$. 
The cluster's gravitational potential $\Phi$, mass density $\rho$, and enclosed mass $M$ are given as functions of radius $r$ by
\begin{subequations}
\begin{align}
    \Phi(r) &= - \frac{G M_{\rm cl}}{\sqrt{r^{2} + b^{2}}}, \\
    \rho(r) &= \frac{3 M_{\rm cl}}{4 \pi b^{3}} \left( 1 + \frac{r^{2}}{b^{2}} \right)^{-5/2}, \\ 
    M(r) &= M_{\rm cl} \left( 1 + \frac{b^{2}}{r^{2}} \right)^{-3/2}.
\end{align} 
The Plummer length $b$ is related to two other convenient reference lengths: 
the \emph{half-mass radius} $r_{\rm h} \equiv [2^{2/3} - 1]^{-1/2} b \approx 1.3 b$ 
and the \emph{virial radius} $r_{\rm v} \equiv 16 b / (3 \pi) \approx 1.7 b$.
The 3D velocity dispersion is given by
\begin{equation}
    \sigma^{2}(r) = - \frac{\Phi(r)}{6}.
\end{equation}
Unless otherwise specified, we evaluate the cluster's properties at $r = r_{\rm h}$ 
and use $r_{\rm v}$ as our default reference length. 
\end{subequations}

Let the cluster contain a population of primordial binaries 
whose semi-major axes $a$ are distributed over many orders of magnitude. 
We choose physically motivated cutoffs for this distribution as follows.  
Let the lower cutoff be given by the binary's near-contact limit during the pre-MS stage, approximated as: 
\begin{equation} \label{eq:Roche_limit_eqmass}
    a_{\rm min} \approx 5 (R_{1} + R_{2}),
\end{equation}
where $R_{1,2}$ are the component radii on the zero-age main sequence.  
We adopt the following approximate ZAMS mass--radius relation \citep[cf.][]{Tout+1996}:
\begin{equation}
    \frac{R(m)}{\RSol} = \left\{ 
    \begin{array}{ll}
        (m/\MSol)^{0.9}, & m \leq \MSol; \\
        (m/\MSol)^{0.6}, & m \geq \MSol.
    \end{array}
    \right.
\end{equation} 
Meanwhile, let the upper cutoff be determined by the tidal (Hill) radius of a circular binary within the cluster potential: 
\begin{equation} \label{eq:amax_radial_plummer}
    a_{\rm max} \sim \left( \frac{m_{\rm b}}{\rho (r)} \right)^{1/3} = b \left( \frac{4 \pi m_{\rm b}}{3 M_{\rm cl}} \right)^{1/3} \left( 1 + \frac{r^{2}}{b^{2}} \right)^{5/6}
\end{equation}
where $m_{\rm b} = m_{1}+m_{2}$ is the mass of a given binary.
At $r = r_{\rm h}$, this yields
\begin{subequations} \label{eq:amax_halfmass_plummer}
\begin{align}
    a_{\rm max} &= 2.17 r_{\rm v} \left( \frac{m_{\rm b}}{M_{\rm cl}} \right)^{1/3} \\
    &= 5600 \AU \left( \frac{r_{\rm v}}{1 \pc} \right) \left( \frac{m_{\rm b}}{\MSol} \right)^{1/3} \left( \frac{M_{\rm cl}}{10^{5} \MSol} \right)^{-1/3}.
\end{align} 
Binaries with separations $\gtrsim a_{\rm max}$ would dissolve within an orbital period simply 
as a result of the cluster's tidal field, even if stellar encounters were somehow suppressed. 
For low-mass stars ($m_{\rm b} \lesssim \MSol$), $a_{\rm max}$ is of the order of the average interparticle distance within a region of locally uniform density.
\end{subequations}

As a binary moves through the cluster, it undergoes numerous weak scattering encounters with background stars, as well as occasional strong encounters. 
The cumulative effect of these encounters is determined by the binary's hardness, 
the ratio of its characteristic orbital velocity to the background velocity dispersion: 
\begin{equation} \label{eq:def_eta}
    \eta = \frac{G (m_{1}+m_{2})}{\sigma^{2} a}.
\end{equation} 
Binaries with $\eta > 1$ and $\eta < 1$ are called hard and soft, respectively.\footnote{Traditionally, 
the hardness of a binary is defined as $\tilde{\eta} = G m_{1} m_{2} / (\bar{m} \sigma^{2} a)$, 
where $\bar{m}$ is the average mass of a background star. 
This definition amounts to a comparison of the dynamical \emph{energy} scales of the binary and third body, rather than their \emph{velocities}. 
Numerical scattering experiments show that the velocity-based definition (Eq.\ \ref{eq:def_eta}, also called the fast/slow boundary) 
better predicts the statistics of binary--single scattering events with arbitrary mass ratios.}
Within a Plummer sphere, the hard/soft boundary is given as a function of radius by setting $\eta = 1$:
\begin{equation} \label{eq:ahs_radial_plummer}
    a_{\rm hs}(r) = \frac{G m_{\rm b}}{\sigma^{2}(r)} = 6(r^{2} + b^{2})^{1/2} \frac{m_{\rm b}}{M_{\rm cl}}.
\end{equation}
At the half-mass radius, we have
\begin{subequations} \label{eq:ahs_halfmass_plummer}
\begin{align}
    a_{\rm hs} &= 5.8 r_{\rm v} \frac{m_{\rm b}}{M_{\rm cl}} \\
    &\approx 12 \AU \left( \frac{r_{\rm v}}{1 \pc} \right) \left( \frac{m_{\rm b}}{\MSol} \right) \left( \frac{M_{\rm cl}}{10^{5} \MSol} \right)^{-1}.
\end{align}
This expression readily generalizes to arbitrary spherical cluster potentials; 
the numerical coefficient varies only by a factor of a few as long as we do not stray too far into the cluster's outer reaches. 
\end{subequations}

It is well understood that, on average, hard binaries harden (i.e., become more tightly bound) as a result of encounters, while soft binaries soften and eventually dissolve \citep{Heggie1975}.
The remainder of this work follows from the realization that, all else being equal, $a_{\rm hs}$ is directly proportional to the binary mass scale $m_{\rm b}$ (which is of the same order as the primary mass $m_{1}$) 
and inversely proportional to the velocity dispsersion $\sigma^{2} \propto M_{\rm cl} / r_{\rm v}$. 
This implies (i) that in a given cluster, massive binaries ($m_{\rm b} \gg \MSol$) are dynamically hard over a much wider range of separations than low-mass binaries ($m_{\rm b} \lesssim \MSol$), 
and likewise (ii) that less compact clusters host more hard binaries of a given mass. 

\begin{figure}
    \centering
    \includegraphics[width=\linewidth]{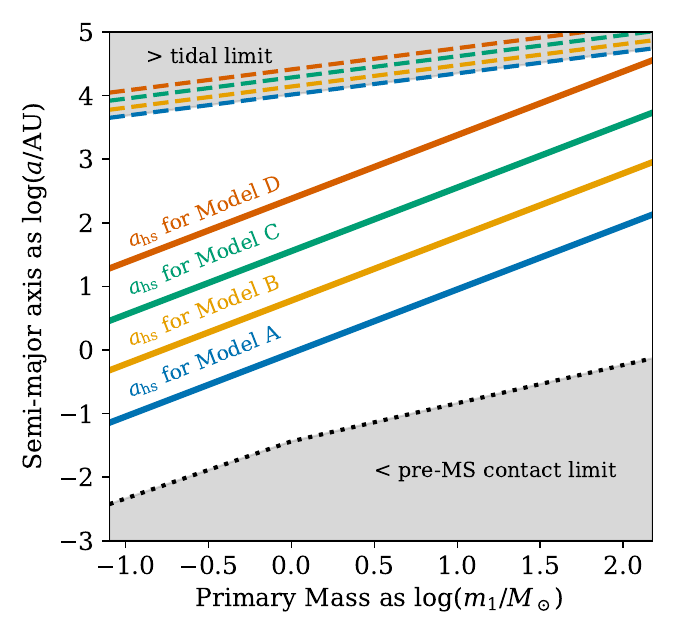}
    \caption{Map of the primary mass--semimajor axis parameter space for binary systems embedded in fiducial Plummer-sphere clusters. 
    Solid lines show the hard/soft boundary as a function of $m_{1}$ in each cluster for binaries with mass ratio $q = 0.5$; 
    the color of each line indicates the cluster parameters described in the text: 
    blue for model A, amber for B, green for C, and red for D. 
    Dashed lines of the corresponding colors shows $a_{\rm max}$ in the same clusters. 
    The dotted black curve shows $a_{\rm min}$. 
    Binaries cannot exist in the shaded regions under our assumptions.}
    \label{fig:logm_loga_map}
\end{figure}

Figure \ref{fig:logm_loga_map} illustrates this point by showing $a_{\rm min}$, $a_{\rm hs}$, and $a_{\rm max}$ as functions of primary mass $m_{1}$ for a sequence of binaries with mass ratio $q = m_{2}/m_{1} = 1/2$. 
We consider four fiducial cluster models with masses $M_{\rm cl} = ( 10, 3 , 1, 0.3 ) \times 10^{5} \MSol$ and $r_{\rm v} = ( 0.5, 1.0, 2.0, 4.0) \pc$, respectively. 
For brevity, we refer to these as models `A' through `D' from greatest density to least. 
Spanning factors of $\approx 300$ in compactness and $\approx 2 \times 10^{4}$ in density, 
these models are intended to be representative for the range of typical GC progenitors 
in that detailed model grids covering a similar range of $M_{\rm cl}$ and $r_{\rm v}$
result in a similar present-day properties to the Milky Way GCs \citep[e.g.,][]{Hong+2018, Kremer2020catalog, Hypki+2025, Giersz+2025}. 

The lower cutoff $a_{\rm min}$ is determined by the ZAMS mass--radius relation and is independent of host cluster properties. 
Plotted in Fig.\ \ref{fig:logm_loga_map} as a dotted black curve, $a_{\rm min}$ ranges from $0.005 \AU$ at the H-burning limit to $0.8 \AU$ at $150 \MSol$. 
The hard/soft boundary $a_{\rm hs}$ (solid curves) and the tidal radius $a_{\rm max}$ (dotted curves) depend 
on both binary and cluster properties, but to differing degrees. 
At a fixed $m_{1}$, $a_{\rm hs}$ spans a factor of $\approx 300$ across the four models in inverse proportion to velocity dispersion (Eq.\ \ref{eq:ahs_radial_plummer}), 
while the tidal radius $a_{\rm max}$ spans a factor of $\approx 25$ due to the scaling with mean cluster density (Eq.\ \ref{eq:amax_radial_plummer}). 
For fixed cluster properties, $a_{\rm hs} \propto m_{\rm b} \sim m_{1}$ increases by a factor of $150/0.08 \approx 1800$ 
between the least and most massive MS stars; 
$a_{\rm max}$ scales as $m_{\rm b}^{1/3} \sim m_{1}^{1/3}$ and thus varies by a comparatively modest factor of $(150/0.08)^{1/3} \approx 12$. 
We conclude that the relative proportions of hard and soft binaries, both within a single cluster and across a diverse population of dense clusters, 
is driven predominantly by the location of the hard/soft boundary; 
$a_{\rm min}$ and $a_{\rm max}$ are comparatively consistent across different environments. 

\subsection{Hard and soft binary fractions} \label{s:Analytics:HardBinaries}

\begin{figure}
    \centering
    \includegraphics[width=\columnwidth]{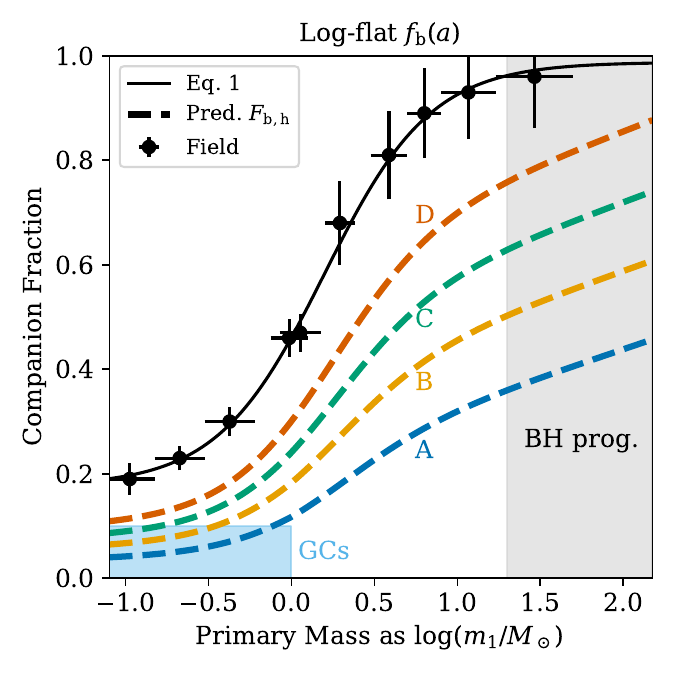}
    \caption{Binary fraction as a function of primary mass. 
    The black points are the measured MF among MS stars in the solar neighborhood
    as compiled in Table 1 of \citet{Offner2023binaries}, 
    with the mass bins and quoted uncertainties shown as horizontal and vertical error bars.
    The black curve is given by Eq.\ \ref{eq:offner_mf_fit}. 
    Dashed colored curves show the predicted primordial hard binary fractions at the half-mass radii 
    of our three fiducial Plummer-sphere clusters, using the same color scheme as in Fig.\ \ref{fig:logm_loga_map}. 
    The shaded regions highlight areas of observational relevance: 
    light blue shows the range of MS stellar masses and observed binary fractions in GCs, 
    while grey shows the approximate mass range for BH progenitor stars.} 
    \label{fig:hardbinfracs}
\end{figure}

\begin{figure}
    \centering
    \includegraphics[width=\columnwidth]{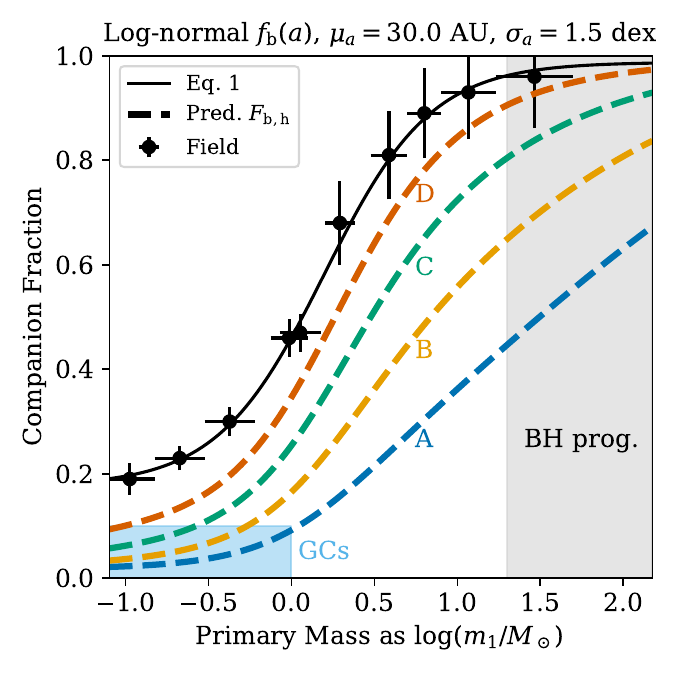}
    \caption{Like Fig.\ \ref{fig:hardbinfracs}, but for a log-normal distribution of semi-major axes 
    with median $\mu_{a} = 30 \AU$ and dispersion $\sigma_{a} = 1.5 \, {\rm dex}$.} 
    \label{fig:hardbinfracs_lognorm}
\end{figure}

Given a model for the primordial binary distribution function and the cluster's initial properties, 
one can compute the hard and soft binary fractions as a function of $m_{\rm b} \sim m_{1}$. 
We now enact a simplified example of such a calculation. 
We consider only single stars and binaries as possible configurations, neglecting triples and higher-order multiplicities. 

For a star of mass $m_{1}$, let the probability of having a companion with semi-major axis between $a$ and $a + \dif a$, 
and whose mass ratio is between $q$ and $q + \dif q$, be
\begin{equation} \label{eq:diff_Fb}
    \dif F_{\rm b} = f_{\rm b}(a, q; m_{1}) \, \dif a \, \dif q, 
\end{equation}
where $f_{\rm b}$ is the IBD.
The binary fraction in any finite region of parameter space is then given by the integral of equation \eqref{eq:diff_Fb} over $a$ and $q$. 
By equation (2), the {\it hard} binary fraction is determined as a function of primary mass by 
\begin{equation}
    F_{\rm b,hard}(m_{1}) = \int_{0}^{1} \int_{a_{\rm min}(m_{1},q)}^{a_{\rm hs}(m_{1},q)} f_{\rm b}(a,q; m_{1}) \, \dif a \, \dif q
\end{equation}
where we set $m_{\rm b} = m_{1} ( 1 + q )$ when evaluating $a_{\rm hs}$. 
The double integral can be evaluated numerically without difficulty for general $f_{\rm b}$. 
The soft binary fraction is given by the analogous integral over $[a_{\rm hs} , a_{\rm max}]$.

Consider now the special case where the mass-ratio dependency is separable and independent of mass, i.e. $f_{\rm b} = f_{\rm b}(a; m_{1}) \xi_{q}(q)$ 
with $\xi_{q}(q)$ a probability density function supported over $0 < q \leq 1$. 
We may then re-write the integral as 
\begin{equation}
    F_{\rm b,h}(m_{\rm 1}) = \int_{a_{\rm min}(\bar{m}_{\rm b})}^{a_{\rm hs}(\bar{m}_{\rm b})} f_{\rm b}(a; m_{1}) \, \dif a, 
\end{equation}
where $\bar{m}_{\rm b} = (1 + \bar{q}) m_{1}$ is the average binary mass at primary mass $m_{1}$, with $\bar{q}$ the average mass ratio. 
Finally, if we assume that the distribution function depends only on mass via an overall normalization $F_{\rm b,tot}(m_{1})$, 
whose value corresponds to the primordial binary fraction at a given mass, 
we have 
\begin{equation}
    F_{\rm b,h}(m_{\rm 1}) = F_{\rm b,tot}(m_{1}) \int_{a_{\rm min}(\bar{m}_{\rm b})}^{a_{\rm hs}(\bar{m}_{\rm b})} f_{\rm b}(a) \, \dif a. 
\end{equation}
As mentioned in the Introduction, the function $F_{\rm b,tot}$ is imprecisely defined 
and is mainly estimated from observations of MS stars in the Galactic field (Eq.\ \ref{eq:offner_mf_fit}). 

\subsubsection{Log-flat separation distribution} \label{s:Analytics:HardBinaries:logflat}

To make further progress, we must specify the distribution function $f_{\rm b}$. 
For the utmost simplicity, let us consider a distribution that is flat in the logarithm of separation, $f_{\rm b} \propto 1/a$ (a.k.a.\ \"{O}pik's law). 
We then have 
\begin{equation}
    F_{\rm b,h}(m_{1}) = F_{\rm b,tot}(m_{1}) \frac{\ln(a_{\rm hs} / a_{\rm min})}{\ln(a_{\rm max}/a_{\rm min})},
\end{equation}
where we suppress the explicit $m_{1}$ dependence inside the logarithmic factors for visual clarity. 
The right-hand side clearly varies smoothly between $0$ and $F_{\rm b,tot}$ for $a_{\rm min} \leq a_{\rm hs} \leq a_{\rm max}$. 
In the limit $a_{\rm hs} \to a_{\rm min}$, we have 
\begin{equation}
    F_{\rm b,h} \simeq \frac{F_{\rm b,tot}}{\ln(a_{\rm max}/a_{\rm min})} \left( \frac{a_{\rm hs}}{a_{\rm min}} - 1 \right).
\end{equation} 
This limit can be realized for low-mass binaries in highly compact clusters such as model A. 

In Figure \ref{fig:hardbinfracs}, we plot $F_{\rm b,h}$ as a function of $m_{1}$ for our three fiducial cluster models, again setting $q = 0.5$ for definiteness. 
In each case, the predicted hard binary fraction in the cluster environment 
is significantly lower than the field value across all stellar masses. 
The more compact the cluster, the smaller the hard binary fraction overall. 
However, as a result of the scaling of $a_{\rm hs}$ with stellar mass, 
the general trend that higher-mass stars occur more frequently in binaries persists. 

It is striking that the predicted $F_{\rm b,h}$ among low-mass stars ($m_{1} \lesssim \MSol$) in dense clusters 
intersects the observationally derived range of close binary fractions in typical GCs ($\lesssim 10 \%$; e.g., \citealt{Milone2012gcbinaryfraction, Ji2015_gcbinaries}). 
The latter is shown by a light-blue-shaded rectangle in the lower-left-hand corner of the plot. 
Meanwhile, in the lower-mass, less-dense clusters (models C and especially D), 
$F_{\rm b,h}$ naturally approaches the primordial value, consistent with the minority of low-density GCs with binary fractions similar to the field.
These bulk estimates miss out on certain nuances, such as the radial variation of the binary fraction within a cluster, but they prove our essential point: 
assuming a universal IBD represented by MS binaries in the solar neighborhood, 
dynamical disruption of soft binaries naturally accounts for observed GC binary fractions 
and predicts their correlations with cluster properties. 

The predicted hard binary fractions of intermediate- ($m_{1} \sim 1 \mbox{--} 8 \MSol$) and high-mass stars ($m_{1} \gtrsim 8 \MSol$), 
while lower than the corresponding field values, 
are significantly larger than observed among low-mass stars in present-day GCs.  
The range of predicted values among the most massive stars is broadly consistent with observational estimates of the close binary fraction for OB stars in YMCs \citep[e.g.,][]{Sana2013_30Dor, Sana2025bloem, Ritchie2022w1obbinaries, Clark2023arches} 
and for the young B stars of the Milky Way's nuclear star cluster \citep{Chu+2023galcentbinaries, Gautam2024galcentbinaries}. 
Previous studies demonstrate that GCs with such high binary fractions among massive stars 
give rise to a host of exotic phenomena, 
such as the seeding of intermediate-mass BHs through stellar collisions \citep[e.g.,][]{Gonzalez2021, GonzalezPrieto2022, GonzalezPrieto2024} 
and the formation of stellar-mass BHs with elevated masses and spins through tidal interactions or accretion \citep[e.g.,][]{Paiella+2025, Kiroglu2025bhaccretion}.
There are also potential implications for the formation of low-mass X-ray binaries, white-dwarf mergers, 
millisecond pulsars, and young magnetars in core-collapsed GCs \citep[e.g.,][]{Ye+2019mspgc, Kremer2020catalog, Kremer+2021frbgc}.

\subsubsection{Log-normal separation distribution} \label{s:Analytics:HardBinaries:lognorm}

In Section \ref{s:Analytics:HardBinaries:logflat}, we computed hard binary fractions assuming that the separations of primordial binaries follow \"{O}pik's law. 
This is a good description of the observed separation distribution among OB stars. 
We now repeat the exercise in an alternative case where $f_{\rm b}$ has a log-normal profile 
with median $\mu_{a} = 30 \AU$ and dispersion $\tilde{\sigma}_{a} = 1.5 \, {\rm dex}$: 
\begin{equation}
    f_{\rm b}(a) = \frac{1}{a \sigma_{a} \sqrt{2 \pi}} \exp\left[ - \frac{\ln^{2}(a/\mu_{a})}{2 \sigma_{a}^{2}} \right],
\end{equation}
where $\sigma_{a} = \tilde{\sigma}_{a} \ln(10)$ is the dispersion in e-folds.
This more closely reflects the period distribution among low-mass field binaries \citep{DM1991, Raghavan+2010, Offner2023binaries}. 
The result may also be expressed in closed form:
\begin{equation}
    F_{\rm b,h}(m_{1}) = F_{\rm b,tot}(m_{1}) \frac{\Psi(a_{\rm hs}/\mu_{a}; \sigma_{a}) - \Psi(a_{\rm min}/\mu_{a} ; \sigma_{a})}{\Psi(a_{\rm max}/\mu_{a} ; \sigma_{a}) - \Psi(a_{\rm min} /\mu_{a} ; \sigma_{a} )}
\end{equation}
where
\begin{equation}
\Psi(x; \sigma) = {\rm erf}\left( \frac{\ln{x}}{\sqrt{2} \sigma} \right)
\end{equation}
is related to the cumulative distribution function of a lognormal distribution with unit median.

Figure \ref{fig:hardbinfracs_lognorm} shows the predicted hard binary fraction for this lognormal separation distribution. 
The predicted hard binary fractions varies with respect to stellar mass qualitatively much as before. 
For a given cluster model, the log-normal distribution produces a smaller hard binary fraction among low-mass stars 
and a higher value among intermediate and massive stars. 
This is due to the changing hierarchy of $\mu_{a}$ and $a_{\rm hs}$ for different masses: 
For $m_{1} \ll \MSol$, we have $a_{\rm hs} \ll \mu_{a}$ (Fig.\ \ref{fig:logm_loga_map}), so most primordial binaries are soft.  
On the other hand, $a_{\rm hs} \gtrsim \mu_{a}$ for $m_{1} \gtrsim \MSol$, so the opposite is true. 

Thus, we conclude that the hard binary fraction tends to be small among low-mass stars in dense clusters 
for a fairly broad class of primordial separation distributions. 
We expect this result to hold as long as the probability density per interval of $\log{a}$ varies slowly across many orders of magnitude, 
such that $a_{\rm hs}$ lies at a low (high) quantile of the distribution for low (high) masses. 

\section{Dynamics of binary-rich clusters} \label{s:Dynamics}

In the previous section, we found that invoking the dynamical disruption of soft binaries naturally explains the observed binary fractions of GCs. 
In this section, we quantify the time and energy required to dissolve a cluster's soft binaries with further analytical estimates. 
We then carry out a detailed numerical experiment using the \texttt{Cluster Monte Carlo} code to corroborate our conclusions. 
Finally, we apply our results to derive constraints on the initial conditions of well-studied Milky Way GCs.

\subsection{Soft binary lifetimes}

Consider a ``target'' binary of mass $m_{\rm b}$ and semi-major axis $a_{\rm b}$ ($\gtrsim a_{\rm hs}$) embedded within a homogeneous stellar background. 
The background consists of single stars of mass $\bar{m}$ and soft binaries of mass $2 \bar{m}$ and typical semi-major axis $\bar{a} \sim \mu_{a}$. 
Let the total number density be $n$, the velocity dispersion $\sigma$, and the binary fraction $F_{\rm b}$. 
The rates at which the target binary experiences strong binary--single and binary--binary encounters can be estimated as follows \citep[e.g.,][]{Fregeau2007}:
\begin{align}
    \Gamma_{\rm bs} &= 4 \sqrt{\pi} \, n_{\rm s} X_{\rm b}^{2} a_{\rm b}^{2} \sigma \left( 1 + \frac{G \max(m_{\rm b}, \bar{m})}{2 \sigma^{2} X_{\rm b} a_{\rm b}} \right), \\ 
    \Gamma_{\rm bb} &= 16 \sqrt{\pi} \, n_{\rm b} X_{\rm b}^{2} A^{2} \sigma \left( 1 + \frac{G \max(m_{\rm b} a_{\rm b}, 2 \bar{m} \bar{a}) }{2 \sigma^{2} X_{\rm b} A^{2}} \right),
\end{align}
where $n_{\rm s} = (1-F_{\rm b}) n$, $n_{\rm b} = F_{\rm b} n$, and $A^{2} = a_{\rm b}^{2} + \bar{a}^{2}$.
The factor $X_{\rm b}$ sets the geometric scattering cross section relative to the size of the orbit; 
we set $X_{\rm b} = 2$ for definiteness. 
The expected interval between strong encounters is then \begin{equation}
    T_{\rm enc} = (\Gamma_{\rm bs} + \Gamma_{\rm bb})^{-1}.
\end{equation} 
When considering soft binaries, binary--binary scattering dominate the strong interaction rate unless $F_{\rm b} \ll 1$ \citep[cf.][]{Hut+1992hardbinaries}. 

To obtain typical values for these quantities for our four fiducial cluster models, 
we estimate the number density as $n = \bar{\rho}_{\rm h} / \bar{m}$, where 
\begin{equation}
    \bar{\rho}_{\rm h} = \frac{3 M_{\rm cl}}{8 \pi r_{\rm h}^{3}} 
\end{equation}
is the average mass density interior to the half-mass radius. 
We take $\bar{m} = 0.6 \MSol$, roughly the average stellar mass under a canonical IMF \citep{Kroupa2001imf, Chabrier2003},
and $F_{\rm b} = 0.3$, the corresponding mean value under our fiducial IBD.

\begin{figure*}
    \centering
    \includegraphics[width=\textwidth]{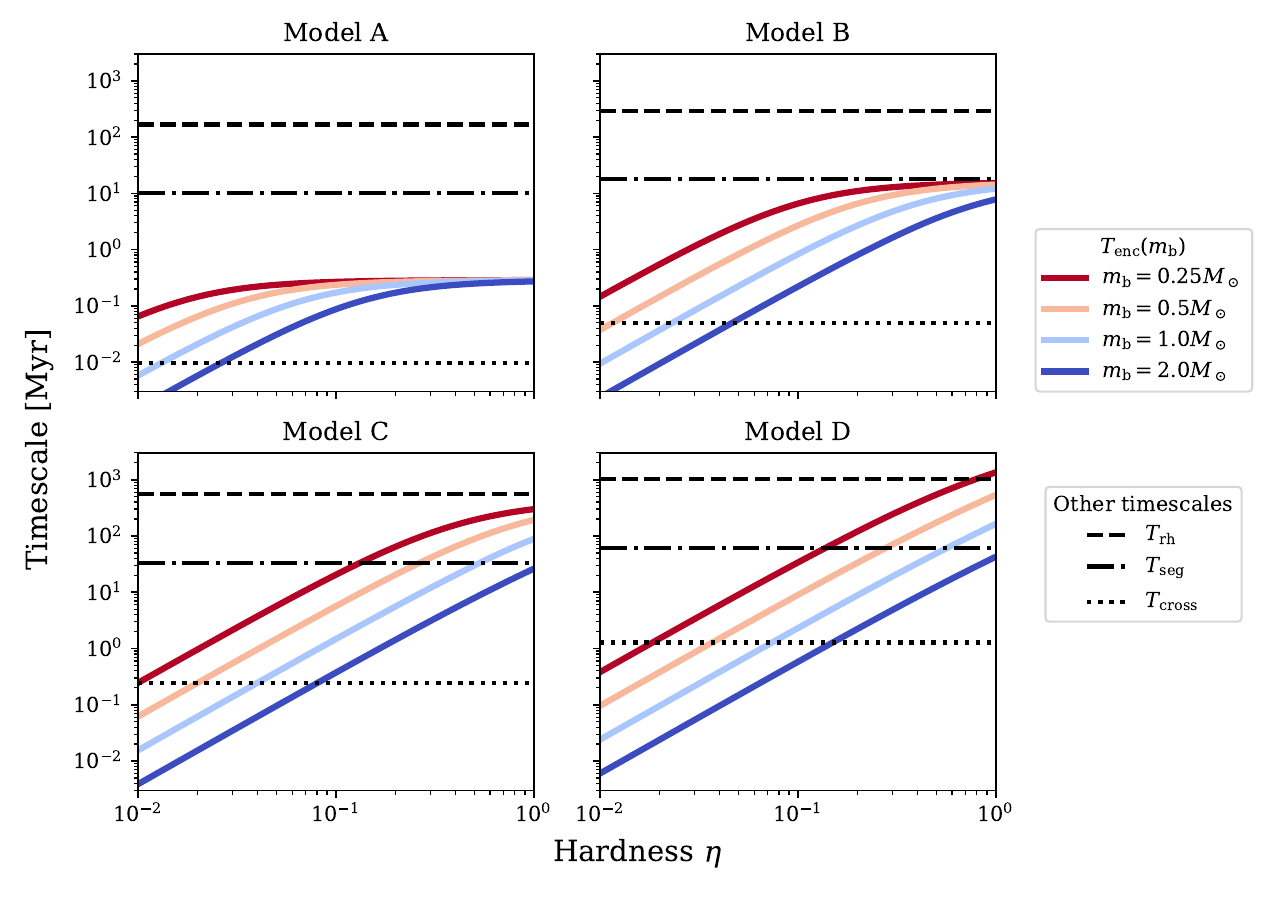}
    \caption{Key timescales related to soft binary evolution within GCs, as computed for fiducial models A through D. 
    The background stars assumed to have an average mass of $\bar{m} = 0.6 \MSol$ and a binary fraction of $F_{\rm b} = 0.3$.
    Colored curves show the strong encounter interval as a function of $\eta$ for soft binaries of various masses. 
    The black horizontal lines indicate the cluster's half-mass crossing time $T_{\rm cross}$, the mass segregation time $T_{\rm seg}$ for $10 \MSol$ BHs, 
    and the half-mass relaxation time $T_{\rm rh}$. 
    Note that the ages of the Milky Way GCs are generally $\gtrsim 10 \Gyr$. 
    \label{fig:softbinarytimescales}}
\end{figure*}

The four panels of Figure \ref{fig:softbinarytimescales} show $T_{\rm enc}$ as a function of $\eta$ for soft binaries in our fiducial cluster models. 
Also shown are three relevant global timescales: 
the half-mass crossing time $T_{\rm cross} = r_{\rm h}/\sigma(r_{\rm h})$ (evaluated at the half-mass radius), the half-mass relaxation time 
\begin{equation}
    T_{\rm rlx} = \frac{0.11 (M_{\rm cl} /\bar{m})}{\ln(0.01 M_{\rm cl}/\bar{m})} T_{\rm cross},
\end{equation}
and the mass-segregation time for massive stars and stellar-mass BHs
\begin{equation}
    T_{\rm seg} \sim \frac{\bar{m}}{10 \MSol} T_{\rm rlx}.
\end{equation}
In the extremely high-density cluster of model A, the large scattering cross-sections of soft binaries bounds $T_{\rm enc}$ strongly from above at $T_{\rm enc} \sim 25 T_{\rm cross}$. 
For solar-mass stars, the widest binaries that can survive longer than a few crossing times lie in the ``semi-soft' regime $\eta \sim 0.1 \mbox{--} 1$. 
In the less-dense models, semi-soft binaries live orders of magnitude longer. 
In model B, their lifetimes are comparable to the mass-segregation timescale of massive objects and BHs; 
in C and D, the lowest-mass soft binaries can last up to a relaxation time. 
Regardless of the typical lifetime, however, the presence of soft binaries at early times is significant for the cluster's long-term dynamical evolution 
because they can participate in the gradual outward transport of mechanical energy within the cluster \citep{Fregeau+2009}. 

\subsection{Energy cost of soft-binary dissolution}

Although soft binaries are short-lived compared to the ages of GCs, 
they can be a dynamically meaningful presence in the early stages of cluster evolution. 
Indeed, using Monte Carlo dynamical models, \citet{Fregeau+2009} found that the rapid dissolution of primordial soft binaries can act as a major energy sink early in a cluster's lifetime. 
To quantify this, we restate a simple analytical estimate of the energy budgets of a cluster's soft and hard binaries made by \citet{Fregeau+2009}. 
Assuming a log-flat separation distribution, the combined gravitational energy of a cluster's soft binaries is
the combined gravitational binding energy of a cluster's soft binaries is
\begin{equation} 
    |E_{\rm b,s}| \sim \frac{F_{\rm b,s}}{\ln{\Lambda_{\rm s}}} \frac{G M_{\rm cl}^{2}}{r_{\rm v}},
\end{equation}
where $\Lambda_{\rm s} \equiv a_{\rm max}/a_{\rm hs} \gg 1$.
Similarly, the energy stored in hard binaries is
\begin{equation}
    |E_{\rm b,h}| \sim \frac{F_{\rm b,h}}{\ln{\Lambda_{\rm h}}} \frac{G M_{\rm cl} \bar{m}}{a_{\rm min}}
\end{equation}
with $\Lambda_{\rm h} = a_{\rm hs}/a_{\rm min} \gg 1$. 
Taking the ratio of these quantities, we have
\begin{align}
    \frac{|E_{\rm b,s}|}{|E_{\rm b,h|}} &= 0.1 \left( \frac{F_{\rm b,s}}{4 F_{\rm b,h}} \frac{\ln{\Lambda_{\rm h}}}{\ln{\Lambda_{\rm s}}} \right) \left( \frac{M_{\rm cl}}{10^{5} \MSol} \right) \nonumber \\
    & \hspace{1cm} \times \left( \frac{r_{\rm v}}{1 \pc} \right)^{-1} \left( \frac{\bar{m}}{\MSol} \right)^{-0.4}.
\end{align}
Evaluating this equation for our fiducial models yields $|E_{\rm b,s}|/|E_{\rm b,h}| \approx 2$ (A), $0.3$ (B), $0.05$ (C), and $0.007$ (D).
For more compact clusters (A, B), soft binary dissolution could, in principle, soak up much of the energy released by the burning of primordial hard binaries, hastening the cluster on its way to core-collapse. 
Whereas a majority of soft, low-mass binaries dissolve on timescales shorter than $T_{\rm rlx}$ (Fig.\ \ref{fig:softbinarytimescales}), 
the timescale of hard binary burning is much longer. 
This mismatch explains why the clusters modeled by \citet{Fregeau+2009} evolve into core-collapsed states on rapid timescales \citep[see also][]{Chatterjee+2013, Kremer2019corecollapse}. 
It also leads us to the slightly counterintuitive conclusion that {\it more-compact} clusters 
are more sensitive to the energy-sink effect of soft binaries \citep[cf.][]{Wang+2022}, 
despite their higher overall energy budgets. 

The discovery of stellar-mass BHs residing within GCs \citep{Maccarone+2007, Irwin+2010, Strader+2012, Chomiuk+2013}, 
alongside the detection of gravitational waves from merging BHs \citep[e.g.,][]{LIGO2016_GW150914, LVK2026_GWTC4}, 
motivated the idea that GCs can retain large populations of BHs, 
whose dynamical burning prevents core collapse for many Gyr after formation 
\citep{Mackey+2007, Mackey+2008, BreenHeggie2013, Morscher2013bhdynamics, Morscher2015bhdynamics, Kremer2019corecollapse, Kremer2020catalog}. 
This is the accepted explanation at present for the fact that most GCs are \emph{not} core collapsed \citep[e.g.,][]{Kremer2019corecollapse, Kremer2020catalog}.
Direct $N$-body studies suggest that the retention of BHs as an additional heat source may offset the heat-sink effect of soft binaries at early times \citep{Wang+2022}. 
This may also be quantified by a straightforward calculation. 

Suppose our model cluster contains $N_{\rm BH}$ BHs, each with mass $m_{\rm \rm BH} \sim 10 \MSol$. 
Due to mass segregation, the BHs form a compact subsystem that evolves more or less independently of the background cluster, 
and its dynamical burning rate is roughly that of a single-species system \citep{BreenHeggie2013}. 
Over its entire evolution from formation to ejection (or merger), each hard binary supplies, on average, a total energy $E_{\rm burn} \sim Q_{\rm burn} G M_{\rm cl} m_{\rm BH} / r_{\rm v}$ to the cluster,
with $Q_{\rm burn} \approx 8$ for equal-mass systems \citep[e.g.,][]{Goodman1984, HeggieHut2003}.
Excluding all other cluster-heating mechanisms, the number of hard BBHs that must be burned to supply the energy $|E_{\rm b,s}|$ required to dissolve of the cluster's soft binaries is
\begin{subequations}
\begin{align}
    N_{\rm burn} &= \frac{|E_{\rm b,s}|}{E_{\rm burn}} = \frac{F_{\rm b,s}}{Q_{\rm burn} \ln{\Lambda_{\rm s}}} \frac{M_{\rm cl}}{m_{\rm BH}} \\ &\approx 8 \left( \frac{F_{\rm b,s}}{0.1} \right) \left( \frac{Q_{\rm burn}}{8} \frac{\ln\Lambda_{\rm s}}{8} \right)^{-1} \nonumber \\
    & \hspace{1.5cm} \times \left( \frac{M_{\rm cl}}{10^{5} \MSol} \right) \left( \frac{m_{\rm BH}}{10 \MSol} \right)^{-1}.
\end{align} 
The total supply of BHs is related to the stellar IMF and the fraction $F_{\rm ret}$ of BHs retained in the cluster after formation. 
Assuming that stellar masses follow a canonical Kroupa IMF \citep{Kroupa2001imf} and that each star with ZAMS mass $\geq 20 \MSol$ forms a BH, 
a cluster produces some $300$ BHs per $10^{5} \MSol$ of initial stellar mass. 
\end{subequations}
Thus, the required {\it fraction} of retained BHs that must be burned to supply an energy $|E_{\rm b,s}|$ is
\begin{subequations}
\begin{align}
    F_{\rm burn} &= \frac{2 N_{\rm burn}}{F_{\rm ret} N_{\rm BH}} \\
    &= 0.24 \left( \frac{F_{\rm b,s}}{0.1} \right) \left( \frac{Q_{\rm burn}}{8} \frac{\ln\Lambda_{\rm s}}{8} \frac{F_{\rm ret}}{0.5} \right)^{-1} \nonumber \\
    & \hspace{1.5cm} \times \left( \frac{\bar{m}}{0.6 \MSol} \right) \left( \frac{m_{\rm BH}}{10 \MSol}\right)^{-1}.
\end{align} 
Only for $F_{\rm b,s} \sim 1$ or $F_{\rm ret} \ll 1$ is a majority of retained BHs potentially consumed in this way. 
It is important to note that the BBHs that participate in burning need not descend directly from primordial stellar binaries, 
since hard BBHs can also be formed in large numbers via three-body interactions over a cluster's lifetime \citep{BreenHeggie2013, Morscher2015bhdynamics, Rodriguez2015, Hong+2018, Kremer2020catalog, Atallah20243bbf, Kremer2026}. 
Our conclusion in this section is therefore robust to theoretical uncertainties regarding the formation of BBHs via isolated binary evolution \citep[e.g.,][]{GallegosGarcia2021bbh}.
\end{subequations} 

Inverting this argument places an upper limit on $F_{\rm b,s}$ in the progenitors of non-core-collapsed GCs (which comprise $\sim 80\%$ of the GC population in the Milky Way). 
If indeed the non-core-collapsed clusters are supported by BH burning, 
all have likely retained $\gtrsim 10$ BHs into the present day \citep{BreenHeggie2013, Askar+2018, Kremer2019corecollapse, Kremer2020catalog, Kremer+2025, Weatherford+2020}. 
By requiring that this number of BHs be retained
after burning has satisfied the energy cost of soft binary dissolution, we find
\begin{align}
    F_{\rm b,s} &\lesssim 0.5 F_{\rm burn,max}  \left( \frac{Q_{\rm burn}}{8} \frac{\ln\Lambda_{\rm s}}{8} \frac{F_{\rm ret}}{0.5} \right) \nonumber \\
    & \hspace{1.5cm} \times \left( \frac{\bar{m}}{0.6 \MSol} \right)^{-1} \left( \frac{m_{\rm BH}}{10 \MSol}\right),
\end{align}
where $F_{\rm burn,max} \sim 1 - [10 / (F_{\rm ret} N_{\rm BH})]$. 
Thus, the fact that most GCs have yet to achieve core collapse disfavors the possibility that the primordial soft binary fraction 
among low-mass stars in GC progenitors was significantly greater than the wide binary fraction among solar-type field stars \citep[cf.][]{Raghavan+2010, Belloni+2017, Moe2019metallicity, ElBadry2019widebinmetallicity}. 
We conclude that BH burning in GCs suffices to dissolve the clusters' primordial soft binaries at early times, 
leaving the hard low-mass binaries largely untouched for many relaxation times. 
The persistence of the BH `engine' is enabled by the continual creation of BBHs via dynamical encounters.

\subsection{{\tt Cluster Monte Carlo} experiment}

To demonstrate that the simple calculations above capture the essence of binary evolution in dense clusters, 
we present a detailed GC simulation conducted with {\tt Cluster Monte Carlo} (\cmc; \citealt{Joshi2000, Pattabiraman2013, Rodriguez2022cmcreview}). 
\cmc\ is an H\'{e}non-type Monte Carlo code \citep{Henon1971montecarlo, Henon1971montecarlo2} that treats all processes necessary to study binaries in dense clusters, 
including direct integration of strong binary--single and binary--binary encounters, 
dynamical binary formation through three-body encounters and tidal capture, 
global two-body relaxation with an external tidal field, and full-lifetime single and binary stellar evolution with {\tt COSMIC}
\citep{Breivik2020cosmic}. 
We adopted largely the same input physics as simulations in the CMC Cluster Catalog \citep{Kremer2020catalog}.  
There are a few differences related to updated stellar evolution prescriptions in {\tt COSMIC}, but none significantly impacts our results. 

We initialized the simulation with $4 \times 10^{5}$ stars distributed according to a King density profile with concentration parameter $W_{0} = 5$ \citep{King1962cluster}. 
We set the cluster's virial radius and metallicity to $r_{\rm v} = 1 \pc$ and $Z = 0.002$, 
and we apply an external tidal field appropriate for a circular orbit $8 \, {\rm kpc}$ in radius within a Milky Way--like potential. 
The cluster's initial mass and radius are similar to those of fiducial model B, albeit with a moderately different density profile. 
All initial stellar masses were sampled from the canonical Kroupa IMF \citep{Kroupa2001imf} between $0.08 \MSol$ and $150 \MSol$, 
and initial binary systems were formed by pairing off stars in such a way as to approximate closely the mass-dependent binary fraction given by equation \eqref{eq:offner_mf_fit} 
with a flat mass-ratio distribution over $0.1 < q < 1$ (or $0.6 < q < 1$ for primary masses above $15 \MSol$) \citep[e.g.,][]{DM1991, Raghavan+2010, Sana+2012}. 
Initial separations were drawn from a log-flat distribution, with lower and upper limits $a_{\rm min}$ (Eq.\ \ref{eq:Roche_limit_eqmass}) and $a_{\rm max}$ (Eq.\ \ref{eq:amax_radial_plummer})
calculated self-consistently for each object according to the binary masses and the local cluster density. 
Initial binary eccentricities were drawn from a thermal distribution (truncated at the Roche limit for close binaries). 

\begin{figure*}
    \centering
    \includegraphics[width=\textwidth]{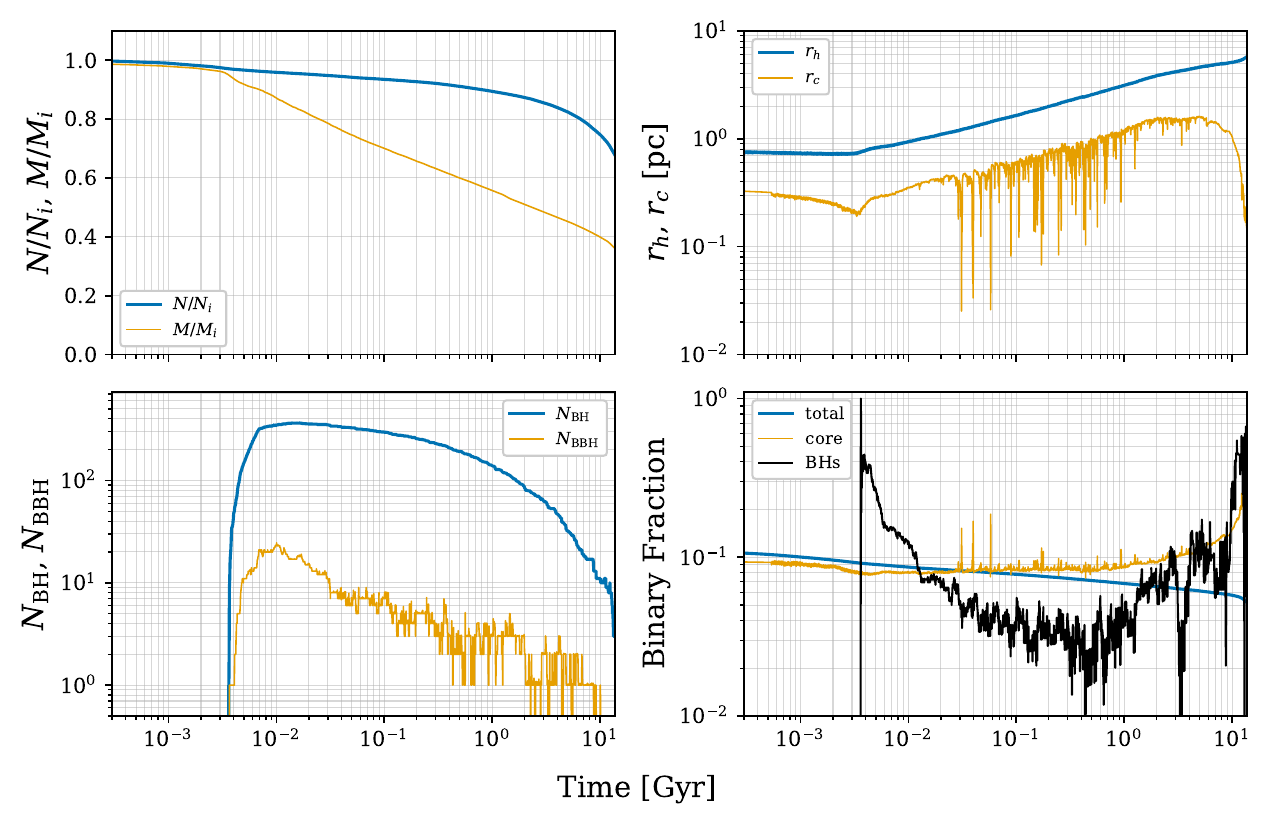}
    \caption{Results of a \cmc\ simulation of a GC with a primordial binary population realized according to our fiducial IBD. 
    Clockwise from the top left, the four panels show (i) the total number of bound stars (blue) and total bound mass (amber), normalized to their initial values; 
    (ii) the half-mass radius (blue) and theoretical (density-weighted) core radius (amber); 
    (iii) the number of bound BHs (blue) and binary BHs (amber); 
    and (iv) the binary fractions of all bound stars (blue), stars in the core (amber), and bound BHs (black). 
    The core and BH data have been smoothed to a moderate degree to enhance visual clarity.}
    \label{fig:cmc_model}
\end{figure*}

Figure \ref{fig:cmc_model} shows the simulation results up to a terminal age of $13.8 \Gyr$. 
The cluster's overall evolution resembles that of models with only hard primordial binaries 
in terms of its total mass, half-mass radius, and BH inventory \citep[cf.][]{Kremer2019corecollapse, Kremer2020catalog}. 
However, the effects of soft binaries can be seen by scrutinizing the evolution of the core radius (amber curve in the second panel) and the total binary fraction (blue curve in the fourth). 
\cmc\ automatically breaks binaries with separations exceeding $10\%$ of the local interparticle distance, extracting the necessary energy from neighboring stars. 
At the same time, the large cross-sections of `semi-soft' binaries ($0.1 \lesssim \eta \lesssim 1$) 
result in numerous calls to the {\tt FEWBODY} module, which handles direct integrations of strong encounters \citep{Fregeau2007}. 
The total binary fraction thus drops from the initial $\sim 30\%$ to $\sim 9\%$ within the first $\sim 3 \Myr$ of evolution; 
this mimics the rapid dissolution rates expected for binaries with $\eta \lesssim 0.1$ (Fig.\ \ref{fig:softbinarytimescales}).\footnote{It 
should be noted that the H\'{e}non Monte Carlo method implemented in \cmc\ is not optimized for clusters with ages less than a relaxation time or so. 
It is nevertheless encouraging that \cmc\ does not produce obviously spurious results at this stage, 
even when stressed by high rates of binary disruption and strong encounters.}
To supply the necessary energy, the cluster core contracts significantly prior to $t \sim 3.5 \Myr$ \citep[cf.][]{Fregeau+2009}. 
The situation abruptly changes at that time, with the onset of supernova mass loss and BH burning; 
these dynamical heating effects quickly restore the core to its initial size and start powering the cluster's long-term expansion. 
On long timescales, the cluster's total binary fraction steadily falls, as the remaining soft binaries dissolve, powered mainly by BH burning, on a relaxation timescale. 
Near the end of the simulation, the cluster exhausts its BH supply and enters core collapse \citep{Chatterjee+2013, Kremer2019corecollapse, Kremer2020catalog}. 
The total binary fraction at late times, at just over $5\%$, is in excellent agreement 
with our estimate of the initial hard binary fraction for the lowest-mass MS stars (cf.\ the `model B' curve in Fig.\ \ref{fig:hardbinfracs}).
The terminal core binary fraction is about $25\%$ (5 times the global value), 
consistent with comparisons between observed core and half-mass binary fractions made by \citet{Milone2012gcbinaryfraction} and \citet{Ji2015_gcbinaries}. 
Thus, the results of our \cmc\ simulation concur with both our analytical predictions and with numerical results produced by other codes \citep{Leigh+2013, Leigh+2015, Hong+2018}. 

Besides confirming that the initial hard binary fraction gives a good estimate of the late-time value, 
our \cmc\ results demonstrate that the dynamical burning of BHs, rather than primordial hard binaries, powers the disruption of soft binaries. 
This extends a previous suggestion by \citet{Wang+2022}, based on a direct $N$-body study of a cluster of $10^{5}$ stars, to more typical GC masses. 
It also underscores BH burning as the dominant process regulating the structure and dynamical evolution of GCs for most of their lifetimes \citep[e.g.,][]{Mackey+2007, Mackey+2008, Morscher2015bhdynamics}, 
with the burning of non-BH binaries only starting when the cluster achieves core collapse \citep{Chatterjee+2013, Kremer2019corecollapse}. 
Outside the core, the contribution of additional dynamical effects to the late-time MS binary fraction, 
such as three-body binary formation and tidal/GW capture, must be small  
(though, again, the situation for compact objects may be very different; \citealt{BreenHeggie2013, Ye+2019mspgc, Ye+2022, Kremer+2021frbgc, Kremer2026}).

\subsection{Application to Milky Way GCs}

\begin{figure*}
    \centering
    \includegraphics[width=0.6\linewidth]{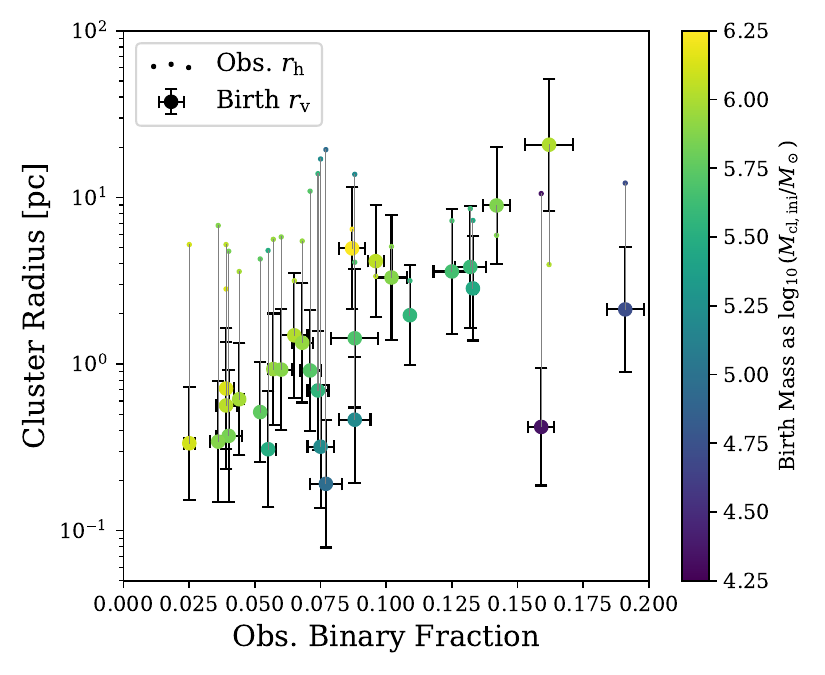}
    \caption{Observed binary fraction versus cluster radius for a sample of 29 well-studied Milky Way GCs \citep{Ji2015_gcbinaries, BaumgardtHilker2018}. 
    Small points show the observed half-mass radius of each cluster, 
    while large points and vertical error bars indicate 
    the range of plausible virial radii implied by our analytical model. 
    The coloration of each point corresponds to the cluster's nominal birth mass as given by \citet{BaumgardtHilker2018}.}
    \label{fig:obs_fbin_vs_rad}
\end{figure*}

Motivated by the good agreement between our analytical formalism and our illustrative \cmc\ result, we apply our formalism to a small sample of well-studied GCs in the Milky Way, 
aiming to extract constraints on their initial conditions from their observed binary fractions \citep[cf.][]{Leigh+2013, Leigh+2015}. 
We select 29 clusters with well-measured binary fractions as reported by \citet{Ji2015_gcbinaries}.
First, we calculate half-mass values of $F_{\rm b,h}$ at a fixed $m_{1} = 0.6 \MSol$ for a sequence of Plummer-sphere models with various half-mass velocity dispersions $\sigma_{\rm h}$ 
(we used a log-flat separation distribution in doing so, but the results are nearly the same for the log-normal distribution).
By setting each cluster's observed binary fraction $F_{\rm b,obs}$ equal to $F_{\rm b,h}$, we invert the relation between $F_{\rm b,h}$ and $\sigma_{\rm h}$ to infer the birth velocity dispersion. 
These values are not, in themselves, especially informative, since this merely constrains the inferred birth cluster to lie on a curve of constant $M_{\rm cl}/r_{\rm v}$. 
To break this degeneracy, we adopt estimates of each cluster's present-day mass $M_{\rm cl,obs}$ and birth mass $M_{\rm cl,ini}$ 
from a catalog of global parameters derived by \citet{BaumgardtHilker2018}. 
Allowing for the true value of $M_{\rm cl,ini}$ to vary by up to a factor of $2$ about the nominal value, 
we constrain each cluster's birth $r_{\rm v}$ to within a factor of $\sim 4$. 

We plot the inferred $r_{\rm v}$ values against $F_{\rm b,obs}$ in Figure \ref{fig:obs_fbin_vs_rad}, color-coding each point according to $M_{\rm cl,ini}$, for the 29 clusters in our sample. 
We also plot each cluster's present-day half-mass radius $r_{\rm h}$, again quoting values from \citet{BaumgardtHilker2018}; 
we connect the two points for each cluster with thin gray lines.
There is considerable scatter among points at a given $F_{\rm b,obs}$, reflecting the variation of cluster birth masses. 
Overall, the inferred birth $r_{\rm v}$ distribution spans $\sim 0.5 \mbox{--} 20 \pc$, in agreement both with 
the observed radii of YMCs in the Milky Way and nearby star-forming galaxies \citep{Bastian+2005, Scheepmaker+2007, ProtegiesZwart2010ymcreview, Krumholz2019starclusters} 
and with the predictions of state-of-the-art hydrodynamical simulations of star formation in massive cloud complexes \citep[e.g.,][]{Grudic+2021, Guszejnov+2022}. 

Importantly, the vast majority of individual clusters pass a ``sanity check'' 
in that their inferred birth $r_{\rm v}$ are smaller than or consistent with their present-day $r_{\rm h}$. 
Only one in our sample, NGC~6121 (a.k.a.\ M4), is significantly smaller today than its inferred birth size, 
meaning that it possesses significantly more binaries than our simple formalism predicts. 
By virtue of its close-in and highly eccentric Galactocentric orbit, 
M4 is thought to have undergone extensive tidal stripping over its lifetime \citep[e.g.,][]{Dinescu+1999}. 
Escaping stars are more likely to be single than binary (a downstream effect of mass segregation; e.g., \citealt{Weatherford+2023}), 
so an enhanced binary fraction would be expected in such cases.
On the whole, our simple estimate based on the hard/soft boundary explains the variation of observed binary fractions within the Milky Way GC population in a concise, coherent manner.

\section{Discussion} \label{s:Discuss}

\subsection{Implications for GC model grids}

Our findings in the previous section, as well as those published in earlier works \citep[e.g.,][]{Leigh+2013, Leigh+2015}, 
suggest that $N$-body simulations of proto-GCs that include only primordial hard binaries \citep[e.g.,][]{Hurley+2007, Kremer2020catalog}
may not fully capture the flow of energy between different reservoirs -- 
soft binaries, hard binaries, and BHs -- at early stages of the cluster's evolution. 
Such models can achieve good approximations of the properties of GCs \emph{in the present day}, when only hard binaries (and perhaps some BHs) remain. 
However, they may be missing certain details about the dynamical processes that affect binary evolution in GCs, 
as well as the production of exotic objects through dynamical interactions \citep[e.g.,][]{Kremer+2018, Kremer+2021frbgc, Ye+2019mspgc, Wang+2022, Kiroglu2025bhaccretion, Wu+2025}. 
It should be noted, however, that we have mainly focused on low-mass binaries in this study. 
In future studies (C.\ E.\ O'Connor et al., in prep.; S.\ Agrawal et al., in prep.), 
we will examine how the sampling of {\it massive} primordial binary systems affects predicted cluster properties and high-energy phenomena.

\subsection{Metallicity and halo binaries}

A potential objection to the idea of a universal IBD is that the metallicity of a stellar population must enter the discussion at some stage. 
The Galactic field binaries on which we base our fiducial IBD mostly have near-solar metallicity, well above that of GC populations. 
Surveys of metal-poor binaries in the Galactic halo reveal clear differences from the metal-rich disk population.
The frequency of close binary companions is observed to be anti-correlated with metallicity; 
concretely, the binary fraction closer than $\sim 100 \AU$ is $\sim 40\%$ among metal-poor ($\lesssim 10 \%$ solar metallicity) FGK dwarfs in the halo 
versus $\sim 25\%$ for solar-metallicity dwarfs in the disk \citep{Moe2019metallicity, ElBadry2019widebinmetallicity}. 
One could therefore argue that we have significantly underestimated the primordial close binary fractions in the metal-poor progenitors of GCs. 

However, it is problematic to take metal-poor binaries in the halo as representative of star formation outcomes in metal-poor environments 
because a poorly determined, but potentially large, fraction of halo stars may be former GC members \citep{Hwang+2021, Phillips+2026}. 
If metal-poor conditions intrinsically enhance the formation of close binary systems (see \citealt{Moe2019metallicity} and references therein), 
then one might expect to find an anticorrelation between 
metallicity and close binary fractions in Milky Way GCs; 
however, no such trend is observed \citep{Milone2012gcbinaryfraction, Ji2015_gcbinaries, BashBelokurov2025}. 
Thus, the differences between field and halo binary populations do not necessarily imply variations of the IBD between metal-rich and metal-poor environments \citep[see also][]{Belloni+2017, Hwang+2021, BashBelokurov2025, Phillips+2026}. 
This is not to say that metallicity has \emph{no} effect on star formation outcomes, 
but rather that potential dynamical processing in parent clusters complicates the interpretation of observations of the halo binary population. 
Further study, encompassing both detailed dynamical models of parent clusters and improved observational constraints on the contribution of disrupted GCs to the halo population, is required to draw firm conclusions. 

\subsection{Triples}

Our overall binary fraction is based on the observed \emph{multiplicity} fraction in the solar neighborhood, rather than the binary fraction per se. 
For low-mass primary stars akin to MS stars in GCs, triples and higher-order multiples 
constitute a small proportion of multi-star systems compared to true binaries \citep{Raghavan+2010, Offner2023binaries}. 
Above $\sim 10 \MSol$, the expected number of companions to each primary star exceeds 2, 
i.e., triples make up the majority. 
Setting aside the challenge of incorporating triples into our so-called universal IBD for the present, 
we remark that triple systems, even when dynamically hard, are expected to be short-lived within a dense cluster, 
whether they be disrupted by dynamical encounters or undergo an inner-binary merger \citep[e.g.,][]{Ivanova+2008,Martinez+2020}. 
The incorporation of triples into detailed cluster models is an ongoing and challenging endeavor, 
but it is unlikely to alter our conclusions about MS binary populations observed in present-day GCs.

\subsection{Comparison with related works}

As we stated at the outset, ours is not the first work to examine how primordial binaries shape the properties of GCs. 
A classic review article by \citet{Hut+1992review} shows that the subject has been well-attended over many decades. 
In this section, we highlight how our work builds on and distills ideas found in previous works by \citet{Ivanova+2005}, \citet{Hurley+2007}, \citet{Fregeau+2009}, \citet{Leigh+2013, Leigh+2015}, and \citet{Wang+2022} on GC dynamics, 
as well as parallel works on other stellar systems.

The work of \citet{Ivanova+2005} is widely cited in discussions of binary evolution in dense clusters, despite its main claim -- 
that a high ($\sim 100\%$) primordial binary fraction is required to explain present-day GC binary fractions -- having been refuted on the basis of direct $N$-body simulations \citep{Hurley+2007} and early \cmc\ models \citep{Fregeau+2009}. 
However, \citet{Hurley+2007} noted that their results and those of \citet{Ivanova+2005} \emph{agree} when only hard binaries are included in the comparison. 
In that narrow sense, our results concur with all three works: Primordial hard binaries are highly resilient over the lifetimes of GCs, 
while soft binaries are mostly short-lived and vanish completely on a relaxation timescale. 
Under our fiducial IBD, we find that the total binary fractions of most GCs are significantly less than their initial values (qualitatively, if not quantitatively, in agreement with \citealt{Ivanova+2005}). 
At the same time, we find that the total primordial binary fractions in typical GCs need not be much greater than $\sim 20\%$ to reproduce present-day observations (in good agreement with \citealt{Hurley+2007}).

The simulations of \citet{Fregeau+2009} also provide a clear example of how binary-rich clusters evolve in the absence of stellar-mass BH subsystems, 
motivating our argument that BH burning is required to avoid an over-abundance of core-collapsed GCs in the present day. 
\citet{Wang+2022} proposed this idea previously based on a direct $N$-body study of a relatively low-mass and diffuse GC ($N=10^{5}$, $r_{\rm v} \sim 2 \pc$). 
We have extended it to more typical GCs with an analytical estimate of the global hard, soft, and BH binary energy budgets over a cluster's lifetime and our illustrative \cmc\ model. 

Meanwhile, \citet{Leigh+2013, Leigh+2015} advanced the idea of using GC binary fractions to constrain their birth densities, 
and concluded that the mass--radius relation for newborn dense clusters should be nearly flat. 
Their results are in qualitative agreement with our analytically inferred birth radii for 29 Milky Way GCs (Fig.\ \ref{fig:obs_fbin_vs_rad}), 
in that clusters across a wide range of masses display a similar spread of birth radii. 

The ideas of binary disruption and mass segregation have been applied to explain stellar multiplicity trends in rich open clusters \citep[e.g.,][]{Geller+2013, Childs+2024, Liu+2025, Wu+2026petarocbinaries}. 
Open clusters are less massive and less compact at birth, and shorter-lived on average, than GC progenitors, 
but they are active enough to exhibit dissolution of wide binaries at early times 
and hardening of close ones in the long term. 
These studies provide further support for the idea that star formation throughout the Milky Way in space and time has produced a universal IBD. 

Finally, a recent work by \citet{Phillips+2026} develops a very similar formalism to ours to forecast the fate of an initial binary population in a cluster based on the hard/soft and tidal boundaries, 
validating its predictions against direct $N$-body integrations of relatively small ($N \sim 10^{4}$) clusters. 
Our work is complementary to theirs in that we deal with retained binary populations rather than ejected binaries in GC stellar streams. 
In so far as the dominant effects shaping the ejected binary population are the disruption of soft binaries and the dynamical burning of BHs, 
Monte Carlo methods suffice to extend the ideas advanced by \citet{Phillips+2026} 
to streams associated with GCs of more typical masses \citep[see, e.g.,][]{Weatherford+2023, Weatherford+2026}. 

\section{Conclusion} \label{s:Summary}

In this study, we have argued for a universal ``initial binary distribution'' (IBD) between the solar neighborhood and Milky Way GCs. 
This inquiry has been motivated by recent advancements in the census of Galactic field binary populations \citep{MoeDiStefano2017, Moe2019metallicity, ElBadry2019widebinmetallicity, Hwang+2021, Offner2023binaries}. 
Our main results are as follows:
\begin{enumerate}
    \item Assuming that the observed demographics of MS binaries in the solar neighborhood is representative of the Milky Way's universal IBD, 
    we correctly predict the observed binary fractions of low-mass MS stars in typical GCs by invoking the disruption of dynamically soft binary systems. 
    Extrapolation to intermediate- and high-mass stars 
    suggests that the hard binary fraction rises with stellar mass, 
    such that a significant fraction of WD progenitors and virtually all NS and BH progenitors in clusters formed in hard binaries. 

    \item By estimating the rate and energy cost of soft binary dissolution as functions of cluster properties, 
    we argue that the dissolution of soft binaries in GCs is powered mainly by dynamical burning of retained stellar-mass BHs. 
    By the same token, soft binaries may also modify the production rates of other high-energy phenomena in GCs. 
   
    \item  We corroborate (1) and (2) with a detailed cluster simulation using the \cmc\ code. 
    After an early burst of wide binary disruption, the total binary fraction declines gradually on a relaxation timescale, powered by BH burning. 
    The late-time binary fraction of our simulated cluster is in excellent agreement with the value predicted by our analytical formalism.

    \item We infer the likely birth radii of a small sample of well-studied Milky Way GCs based on their observed binary fractions. 
    The inferred values are $\sim 0.5 \mbox{--} 20 \pc$, consistent with the properties of YMCs and SSCs in the local universe. 
\end{enumerate}

Remaining uncertainties include how metallicity shapes primordial binary demographics; 
how the flow of energy between hard binaries, soft binaries, and stellar-mass BHs differs within clusters of different masses and sizes; 
and how to incorporate triple and higher-multiple systems into simulations. 
These motivate further numerical studies of primordial binary evolution and survival within GCs and other dense clusters, 
as well as their downstream effects on other in-cluster phenomena. 
We conclude that the inclusion of a complete, observationally motivated IBD in modern GC models 
is essential to understanding these singularly dynamic environments. 

\begin{acknowledgments}
This work was supported by NSF Grants AST-2108624 and AST-2511543 at Northwestern University. 
C.E.O.\ thanks Tri Nguyen and Alex Ji for helpful discussions 
and acknowledges financial support from a CIERA Postdoctoral Fellowship. 
This work used computing resources at the Quest high-performance computing facility provided by CIERA under NSF Grant PHY-2406802.
Quest is jointly supported by Northwestern University's Office of the Provost, the Office for Research, and Northwestern University Information Technology.

\software{{\tt Cluster Monte Carlo} \citep{Joshi2000, Pattabiraman2013, Rodriguez2022cmcreview}, {\tt COSMIC} \citep{Breivik2020cosmic}, {\tt Matplotlib} \citep{Hunter2007matplotlib}, {\tt NumPy} \citep{Harris+2020_NumPy}}
\end{acknowledgments}

\bibliography{ms_GCbinaries}

\begin{thebibliography}{}
\expandafter\ifx\csname natexlab\endcsname\relax\def\natexlab#1{#1}\fi
\providecommand{\url}[1]{\href{#1}{#1}}
\providecommand{\dodoi}[1]{doi:~\href{http://doi.org/#1}{\nolinkurl{#1}}}
\providecommand{\doeprint}[1]{\href{http://ascl.net/#1}{\nolinkurl{http://ascl.net/#1}}}
\providecommand{\doarXiv}[1]{\href{https://arxiv.org/abs/#1}{\nolinkurl{https://arxiv.org/abs/#1}}}

\bibitem[{B.~P. {Abbott} {et~al.}(2016){Abbott}, {Abbott}, {Abbott}, {Abernathy}, {Acernese}, {Ackley}, {Adams}, {Adams}, {Addesso}, {Adhikari}, {Adya}, {Affeldt}, {Agathos}, {Agatsuma}, {Aggarwal}, {Aguiar}, {Aiello}, {Ain}, {Ajith}, {Allen}, {Allocca}, {Altin}, {Anderson}, {Anderson}, {Arai}, {Araya}, {Arceneaux}, {Areeda}, {Arnaud}, {Arun}, {Ascenzi}, {Ashton}, {Ast}, {Aston}, {Astone}, {Aufmuth}, {Aulbert}, {Babak}, {Bacon}, {Bader}, {Baker}, {Baldaccini}, {Ballardin}, {Ballmer}, {Barayoga}, {Barclay}, {Barish}, {Barker}, {Barone}, {Barr}, {Barsotti}, {Barsuglia}, {Barta}, {Bartlett}, {Bartos}, {Bassiri}, {Basti}, {Batch}, {Baune}, {Bavigadda}, {Bazzan}, {Behnke}, {Bejger}, {Bell}, {Bell}, {Berger}, {Bergman}, {Bergmann}, {Berry}, {Bersanetti}, {Bertolini}, {Betzwieser}, {Bhagwat}, {Bhandare}, {Bilenko}, {Billingsley}, {Birch}, {Birney}, {Biscans}, {Bisht}, {Bitossi}, {Biwer}, {Bizouard}, {Blackburn}, {Blair}, {Blair}, {Blair}, {Bloemen}, {Bock}, {Bodiya}, {Boer}, {Bogaert}, {Bogan}, {Bohe},
  {Boh{\'e}mier}, {Bojtos}, {Bond}, {Bondu}, {Bonnand}, {Boom}, {Bork}, {Boschi}, {Bose}, {Bouffanais}, {Bozzi}, {Bradaschia}, {Brady}, {Braginsky}, {Branchesi}, {Brau}, {Briant}, {Brillet}, {Brinkmann}, {Brisson}, {Brockill}, {Brooks}, {Brown}, {Brown}, {Brown}, {Buchanan}, {Buikema}, {Bulik}, {Bulten}, {Buonanno}, {Buskulic}, {Buy}, {Byer}, {Cabero}, {Cadonati}, {Cagnoli}, {Cahillane}, {Calder{\'o}n Bustillo}, {Callister}, {Calloni}, {Camp}, {Cannon}, {Cao}, {Capano}, {Capocasa}, {Carbognani}, {Caride}, {Casanueva Diaz}, {Casentini}, {Caudill}, {Cavagli{\`a}}, {Cavalier}, {Cavalieri}, {Cella}, {Cepeda}, {Cerboni Baiardi}, {Cerretani}, {Cesarini}, {Chakraborty}, {Chalermsongsak}, {Chamberlin}, {Chan}, {Chao}, {Charlton}, {Chassande-Mottin}, {Chen}, {Chen}, {Cheng}, {Chincarini}, {Chiummo}, {Cho}, {Cho}, {Chow}, {Christensen}, {Chu}, {Chua}, {Chung}, {Ciani}, {Clara}, {Clark}, {Clayton}, {Cleva}, {Coccia}, {Cohadon}, {Cokelaer}, {Colla}, {Collette}, {Cominsky}, {Constancio}, {Conte}, {Conti}, {Cook},
  {Corbitt}, {Cornish}, {Corsi}, {Cortese}, {Costa}, {Coughlin}, {Coughlin}, {Coulon}, {Countryman}, {Couvares}, {Cowan}, {Coward}, {Cowart}, \& {Coyne}}]{LIGO2016_GW150914}
{Abbott}, B.~P., {Abbott}, R., {Abbott}, T.~D., {et~al.} 2016, \bibinfo{title}{{GW150914: First results from the search for binary black hole coalescence with Advanced LIGO},} \prd, 93, 122003, \dodoi{10.1103/PhysRevD.93.122003}

\bibitem[{M.~D. {Albrow} {et~al.}(2001){Albrow}, {Gilliland}, {Brown}, {Edmonds}, {Guhathakurta}, \& {Sarajedini}}]{Albrow2001_47tucbinaries}
{Albrow}, M.~D., {Gilliland}, R.~L., {Brown}, T.~M., {et~al.} 2001, \bibinfo{title}{{The Frequency of Binary Stars in the Core of 47 Tucanae},} \apj, 559, 1060, \dodoi{10.1086/322353}

\bibitem[{A. {Askar} {et~al.}(2018){Askar}, {Arca Sedda}, \& {Giersz}}]{Askar+2018}
{Askar}, A., {Arca Sedda}, M., \& {Giersz}, M. 2018, \bibinfo{title}{{MOCCA-SURVEY Database I: Galactic globular clusters harbouring a black hole subsystem},} \mnras, 478, 1844, \dodoi{10.1093/mnras/sty1186}

\bibitem[{D. {Atallah} {et~al.}(2024){Atallah}, {Weatherford}, {Trani}, \& {Rasio}}]{Atallah20243bbf}
{Atallah}, D., {Weatherford}, N.~C., {Trani}, A.~A., \& {Rasio}, F.~A. 2024, \bibinfo{title}{{On Binary Formation from Three Initially Unbound Bodies},} \apj, 970, 112, \dodoi{10.3847/1538-4357/ad5185}

\bibitem[{H. {Bartko} {et~al.}(2010){Bartko}, {Martins}, {Trippe}, {Fritz}, {Genzel}, {Ott}, {Eisenhauer}, {Gillessen}, {Paumard}, {Alexander}, {Dodds-Eden}, {Gerhard}, {Levin}, {Mascetti}, {Nayakshin}, {Perets}, {Perrin}, {Pfuhl}, {Reid}, {Rouan}, {Zilka}, \& {Sternberg}}]{Bartko+2010}
{Bartko}, H., {Martins}, F., {Trippe}, S., {et~al.} 2010, \bibinfo{title}{{An Extremely Top-Heavy Initial Mass Function in the Galactic Center Stellar Disks},} \apj, 708, 834, \dodoi{10.1088/0004-637X/708/1/834}

\bibitem[{D. {Bashi} \& V. {Belokurov}(2025){Bashi} \& {Belokurov}}]{BashBelokurov2025}
{Bashi}, D., \& {Belokurov}, V. 2025, \bibinfo{title}{{Fewer companions in the crowd: the low close binary fraction in globular clusters from Gaia RVS},} \mnras, 541, 2008, \dodoi{10.1093/mnras/staf1102}

\bibitem[{N. {Bastian} {et~al.}(2005){Bastian}, {Gieles}, {Lamers}, {Scheepmaker}, \& {de Grijs}}]{Bastian+2005}
{Bastian}, N., {Gieles}, M., {Lamers}, H.~J.~G.~L.~M., {Scheepmaker}, R.~A., \& {de Grijs}, R. 2005, \bibinfo{title}{{The star cluster population of M 51. II. Age distribution and relations among the derived parameters},} \aap, 431, 905, \dodoi{10.1051/0004-6361:20041078}

\bibitem[{H. {Baumgardt} \& M. {Hilker}(2018){Baumgardt} \& {Hilker}}]{BaumgardtHilker2018}
{Baumgardt}, H., \& {Hilker}, M. 2018, \bibinfo{title}{{A catalogue of masses, structural parameters, and velocity dispersion profiles of 112 Milky Way globular clusters},} \mnras, 478, 1520, \dodoi{10.1093/mnras/sty1057}

\bibitem[{D. {Belloni} {et~al.}(2017){Belloni}, {Askar}, {Giersz}, {Kroupa}, \& {Rocha-Pinto}}]{Belloni+2017}
{Belloni}, D., {Askar}, A., {Giersz}, M., {Kroupa}, P., \& {Rocha-Pinto}, H.~J. 2017, \bibinfo{title}{{On the initial binary population for star cluster simulations},} \mnras, 471, 2812, \dodoi{10.1093/mnras/stx1763}

\bibitem[{P.~G. {Breen} \& D.~C. {Heggie}(2013){Breen} \& {Heggie}}]{BreenHeggie2013}
{Breen}, P.~G., \& {Heggie}, D.~C. 2013, \bibinfo{title}{{Dynamical evolution of black hole subsystems in idealized star clusters},} \mnras, 432, 2779, \dodoi{10.1093/mnras/stt628}

\bibitem[{K. {Breivik} {et~al.}(2020){Breivik}, {Coughlin}, {Zevin}, {Rodriguez}, {Kremer}, {Ye}, {Andrews}, {Kurkowski}, {Digman}, {Larson}, \& {Rasio}}]{Breivik2020cosmic}
{Breivik}, K., {Coughlin}, S., {Zevin}, M., {et~al.} 2020, \bibinfo{title}{{COSMIC Variance in Binary Population Synthesis},} \apj, 898, 71, \dodoi{10.3847/1538-4357/ab9d85}

\bibitem[{G. {Chabrier}(2003){Chabrier}}]{Chabrier2003}
{Chabrier}, G. 2003, \bibinfo{title}{{Galactic Stellar and Substellar Initial Mass Function},} \pasp, 115, 763, \dodoi{10.1086/376392}

\bibitem[{S. {Chatterjee} {et~al.}(2013){Chatterjee}, {Umbreit}, {Fregeau}, \& {Rasio}}]{Chatterjee+2013}
{Chatterjee}, S., {Umbreit}, S., {Fregeau}, J.~M., \& {Rasio}, F.~A. 2013, \bibinfo{title}{{Understanding the dynamical state of globular clusters: core-collapsed versus non-core-collapsed},} \mnras, 429, 2881, \dodoi{10.1093/mnras/sts464}

\bibitem[{A.~C. {Childs} {et~al.}(2024){Childs}, {Geller}, {von Hippel}, {Motherway}, \& {Zwicker}}]{Childs+2024}
{Childs}, A.~C., {Geller}, A.~M., {von Hippel}, T., {Motherway}, E., \& {Zwicker}, C. 2024, \bibinfo{title}{{Goodbye to Chi by Eye: A Bayesian Analysis of Photometric Binaries in Six Open Clusters},} \apj, 962, 41, \dodoi{10.3847/1538-4357/ad18c0}

\bibitem[{L. {Chomiuk} {et~al.}(2013){Chomiuk}, {Strader}, {Maccarone}, {Miller-Jones}, {Heinke}, {Noyola}, {Seth}, \& {Ransom}}]{Chomiuk+2013}
{Chomiuk}, L., {Strader}, J., {Maccarone}, T.~J., {et~al.} 2013, \bibinfo{title}{{A Radio-selected Black Hole X-Ray Binary Candidate in the Milky Way Globular Cluster M62},} \apj, 777, 69, \dodoi{10.1088/0004-637X/777/1/69}

\bibitem[{D.~S. {Chu} {et~al.}(2023){Chu}, {Do}, {Ghez}, {Gautam}, {Ciurlo}, {O'neil}, {Hosek}, {Hees}, {Naoz}, {Sakai}, {Lu}, {Chen}, {Bentley}, {Becklin}, \& {Matthews}}]{Chu+2023galcentbinaries}
{Chu}, D.~S., {Do}, T., {Ghez}, A., {et~al.} 2023, \bibinfo{title}{{Evidence of a Decreased Binary Fraction for Massive Stars within 20 milliparsecs of the Supermassive Black Hole at the Galactic Center},} \apj, 948, 94, \dodoi{10.3847/1538-4357/acc93e}

\bibitem[{J.~S. {Clark} {et~al.}(2023){Clark}, {Lohr}, {Najarro}, {Patrick}, \& {Ritchie}}]{Clark2023arches}
{Clark}, J.~S., {Lohr}, M.~E., {Najarro}, F., {Patrick}, L.~R., \& {Ritchie}, B.~W. 2023, \bibinfo{title}{{The Arches cluster revisited - IV. Observational constraints on the binary properties of very massive stars},} \mnras, 521, 4473, \dodoi{10.1093/mnras/stad449}

\bibitem[{J.~S. {Clark} {et~al.}(2020){Clark}, {Ritchie}, \& {Negueruela}}]{Clark+2020}
{Clark}, J.~S., {Ritchie}, B.~W., \& {Negueruela}, I. 2020, \bibinfo{title}{{A VLT/FLAMES survey for massive binaries in Westerlund 1. VII. Cluster census},} \aap, 635, A187, \dodoi{10.1051/0004-6361/201935903}

\bibitem[{D.~I. {Dinescu} {et~al.}(1999){Dinescu}, {Girard}, \& {van Altena}}]{Dinescu+1999}
{Dinescu}, D.~I., {Girard}, T.~M., \& {van Altena}, W.~F. 1999, \bibinfo{title}{{Space Velocities of Globular Clusters. III. Cluster Orbits and Halo Substructure},} \aj, 117, 1792, \dodoi{10.1086/300807}

\bibitem[{A. {Duquennoy} \& M. {Mayor}(1991){Duquennoy} \& {Mayor}}]{DM1991}
{Duquennoy}, A., \& {Mayor}, M. 1991, \bibinfo{title}{{Multiplicity among Solar Type Stars in the Solar Neighbourhood - Part Two - Distribution of the Orbital Elements in an Unbiased Sample},} \aap, 248, 485

\bibitem[{K. {El-Badry} \& H.-W. {Rix}(2019){El-Badry} \& {Rix}}]{ElBadry2019widebinmetallicity}
{El-Badry}, K., \& {Rix}, H.-W. 2019, \bibinfo{title}{{The wide binary fraction of solar-type stars: emergence of metallicity dependence at a < 200 au},} \mnras, 482, L139, \dodoi{10.1093/mnrasl/sly206}

\bibitem[{A.~C. {Fabian} {et~al.}(1975){Fabian}, {Pringle}, \& {Rees}}]{Fabian+1975}
{Fabian}, A.~C., {Pringle}, J.~E., \& {Rees}, M.~J. 1975, \bibinfo{title}{{Tidal capture formation of binary systems and X-ray sources in globular clusters.},} \mnras, 172, 15, \dodoi{10.1093/mnras/172.1.15P}

\bibitem[{J.~M. {Fregeau} {et~al.}(2009){Fregeau}, {Ivanova}, \& {Rasio}}]{Fregeau+2009}
{Fregeau}, J.~M., {Ivanova}, N., \& {Rasio}, F.~A. 2009, \bibinfo{title}{{Evolution of the Binary Fraction in Dense Stellar Systems},} \apj, 707, 1533, \dodoi{10.1088/0004-637X/707/2/1533}

\bibitem[{J.~M. {Fregeau} \& F.~A. {Rasio}(2007){Fregeau} \& {Rasio}}]{Fregeau2007}
{Fregeau}, J.~M., \& {Rasio}, F.~A. 2007, \bibinfo{title}{{Monte Carlo Simulations of Globular Cluster Evolution. IV. Direct Integration of Strong Interactions},} \apj, 658, 1047, \dodoi{10.1086/511809}

\bibitem[{M. {Gallegos-Garcia} {et~al.}(2021){Gallegos-Garcia}, {Berry}, {Marchant}, \& {Kalogera}}]{GallegosGarcia2021bbh}
{Gallegos-Garcia}, M., {Berry}, C. P.~L., {Marchant}, P., \& {Kalogera}, V. 2021, \bibinfo{title}{{Binary Black Hole Formation with Detailed Modeling: Stable Mass Transfer Leads to Lower Merger Rates},} \apj, 922, 110, \dodoi{10.3847/1538-4357/ac2610}

\bibitem[{A.~K. {Gautam} {et~al.}(2024){Gautam}, {Do}, {Ghez}, {Chu}, {Hosek}, {Sakai}, {Naoz}, {Morris}, {Ciurlo}, {Haggard}, \& {Lu}}]{Gautam2024galcentbinaries}
{Gautam}, A.~K., {Do}, T., {Ghez}, A.~M., {et~al.} 2024, \bibinfo{title}{{An Estimate of the Binary Star Fraction among Young Stars at the Galactic Center: Possible Evidence of a Radial Dependence},} \apj, 964, 164, \dodoi{10.3847/1538-4357/ad26e6}

\bibitem[{A.~M. {Geller} {et~al.}(2013){Geller}, {de Grijs}, {Li}, \& {Hurley}}]{Geller+2013}
{Geller}, A.~M., {de Grijs}, R., {Li}, C., \& {Hurley}, J.~R. 2013, \bibinfo{title}{{Consequences of Dynamical Disruption and Mass Segregation for the Binary Frequencies of Star Clusters},} \apj, 779, 30, \dodoi{10.1088/0004-637X/779/1/30}

\bibitem[{M. {Giersz} {et~al.}(2025){Giersz}, {Askar}, {Hypki}, {Hong}, {Wiktorowicz}, \& {Hellstrom}}]{Giersz+2025}
{Giersz}, M., {Askar}, A., {Hypki}, A., {et~al.} 2025, \bibinfo{title}{{Simulations of Globular Cluster Evolution with Multiple Stellar Populations},} arXiv e-prints, arXiv:2510.06942, \dodoi{10.48550/arXiv.2510.06942}

\bibitem[{E. {Gonz{\'a}lez Prieto} {et~al.}(2021){Gonz{\'a}lez Prieto}, {Kremer}, {Chatterjee}, {Fragione}, {Rodriguez}, {Weatherford}, {Ye}, \& {Rasio}}]{Gonzalez2021}
{Gonz{\'a}lez Prieto}, E., {Kremer}, K., {Chatterjee}, S., {et~al.} 2021, \bibinfo{title}{{Intermediate-mass Black Holes from High Massive-star Binary Fractions in Young Star Clusters},} \apjl, 908, L29, \dodoi{10.3847/2041-8213/abdf5b}

\bibitem[{E. {Gonz{\'a}lez Prieto} {et~al.}(2022){Gonz{\'a}lez Prieto}, {Kremer}, {Fragione}, {Martinez}, {Weatherford}, {Zevin}, \& {Rasio}}]{GonzalezPrieto2022}
{Gonz{\'a}lez Prieto}, E., {Kremer}, K., {Fragione}, G., {et~al.} 2022, \bibinfo{title}{{Intermediate-mass Black Holes on the Run from Young Star Clusters},} \apj, 940, 131, \dodoi{10.3847/1538-4357/ac9b0f}

\bibitem[{E. {Gonz{\'a}lez Prieto} {et~al.}(2024){Gonz{\'a}lez Prieto}, {Weatherford}, {Fragione}, {Kremer}, \& {Rasio}}]{GonzalezPrieto2024}
{Gonz{\'a}lez Prieto}, E., {Weatherford}, N.~C., {Fragione}, G., {Kremer}, K., \& {Rasio}, F.~A. 2024, \bibinfo{title}{{Intermediate-mass Black Hole Progenitors from Stellar Collisions in Dense Star Clusters},} \apj, 969, 29, \dodoi{10.3847/1538-4357/ad43d6}

\bibitem[{J. {Goodman}(1984){Goodman}}]{Goodman1984}
{Goodman}, J. 1984, \bibinfo{title}{{Homologous evolution of stellar systems after core collapse},} \apj, 280, 298, \dodoi{10.1086/161996}

\bibitem[{M.~Y. {Grudi{\'c}} {et~al.}(2021){Grudi{\'c}}, {Guszejnov}, {Hopkins}, {Offner}, \& {Faucher-Gigu{\`e}re}}]{Grudic+2021}
{Grudi{\'c}}, M.~Y., {Guszejnov}, D., {Hopkins}, P.~F., {Offner}, S. S.~R., \& {Faucher-Gigu{\`e}re}, C.-A. 2021, \bibinfo{title}{{STARFORGE: Towards a comprehensive numerical model of star cluster formation and feedback},} \mnras, 506, 2199, \dodoi{10.1093/mnras/stab1347}

\bibitem[{D. {Guszejnov} {et~al.}(2022){Guszejnov}, {Markey}, {Offner}, {Grudi{\'c}}, {Faucher-Gigu{\`e}re}, {Rosen}, \& {Hopkins}}]{Guszejnov+2022}
{Guszejnov}, D., {Markey}, C., {Offner}, S. S.~R., {et~al.} 2022, \bibinfo{title}{{Cluster assembly and the origin of mass segregation in the STARFORGE simulations},} \mnras, 515, 167, \dodoi{10.1093/mnras/stac1737}

\bibitem[{D. {Guszejnov} {et~al.}(2023){Guszejnov}, {Raju}, {Offner}, {Grudi{\'c}}, {Faucher-Gigu{\`e}re}, {Hopkins}, \& {Rosen}}]{Guszejnov+2023}
{Guszejnov}, D., {Raju}, A.~N., {Offner}, S. S.~R., {et~al.} 2023, \bibinfo{title}{{Effects of the environment on the multiplicity properties of stars in the STARFORGE simulations},} \mnras, 518, 4693, \dodoi{10.1093/mnras/stac3268}

\bibitem[{C.~R. {Harris} {et~al.}(2020){Harris}, {Millman}, {van der Walt}, {Gommers}, {Virtanen}, {Cournapeau}, {Wieser}, {Taylor}, {Berg}, {Smith}, {Kern}, {Picus}, {Hoyer}, {van Kerkwijk}, {Brett}, {Haldane}, {del R{\'\i}o}, {Wiebe}, {Peterson}, {G{\'e}rard-Marchant}, {Sheppard}, {Reddy}, {Weckesser}, {Abbasi}, {Gohlke}, \& {Oliphant}}]{Harris+2020_NumPy}
{Harris}, C.~R., {Millman}, K.~J., {van der Walt}, S.~J., {et~al.} 2020, \bibinfo{title}{{Array programming with NumPy},} \nat, 585, 357, \dodoi{10.1038/s41586-020-2649-2}

\bibitem[{D. {Heggie} \& P. {Hut}(2003){Heggie} \& {Hut}}]{HeggieHut2003}
{Heggie}, D., \& {Hut}, P. 2003, {The Gravitational Million-Body Problem: A Multidisciplinary Approach to Star Cluster Dynamics}

\bibitem[{D.~C. {Heggie}(1975){Heggie}}]{Heggie1975}
{Heggie}, D.~C. 1975, \bibinfo{title}{{Binary evolution in stellar dynamics.},} \mnras, 173, 729, \dodoi{10.1093/mnras/173.3.729}

\bibitem[{M. {H{\'e}non}(1971{\natexlab{a}}){H{\'e}non}}]{Henon1971montecarlo}
{H{\'e}non}, M. 1971{\natexlab{a}}, \bibinfo{title}{{Monte Carlo Models of Star Clusters (Part of the Proceedings of the IAU Colloquium No. 10, held in Cambridge, England, August 12-15, 1970.)},} \apss, 13, 284, \dodoi{10.1007/BF00649159}

\bibitem[{M. {H{\'e}non}(1971{\natexlab{b}}){H{\'e}non}}]{Henon1971montecarlo2}
{H{\'e}non}, M. 1971{\natexlab{b}}, \bibinfo{title}{{The Monte Carlo Method (Papers appear in the Proceedings of IAU Colloquium No. 10 Gravitational N-Body Problem (ed. by Myron Lecar), R. Reidel Publ. Co. , Dordrecht-Holland.)},} \apss, 14, 151, \dodoi{10.1007/BF00649201}

\bibitem[{J. {Hong} {et~al.}(2018){Hong}, {Vesperini}, {Askar}, {Giersz}, {Szkudlarek}, \& {Bulik}}]{Hong+2018}
{Hong}, J., {Vesperini}, E., {Askar}, A., {et~al.} 2018, \bibinfo{title}{{Binary black hole mergers from globular clusters: the impact of globular cluster properties},} \mnras, 480, 5645, \dodoi{10.1093/mnras/sty2211}

\bibitem[{J. {Hong} {et~al.}(2017){Hong}, {Vesperini}, {Belloni}, \& {Giersz}}]{Hong+2017}
{Hong}, J., {Vesperini}, E., {Belloni}, D., \& {Giersz}, M. 2017, \bibinfo{title}{{Dynamical formation of cataclysmic variables in globular clusters},} \mnras, 464, 2511, \dodoi{10.1093/mnras/stw2595}

\bibitem[{J. {Hong} {et~al.}(2015){Hong}, {Vesperini}, {Sollima}, {McMillan}, {D'Antona}, \& {D'Ercole}}]{Hong+2015}
{Hong}, J., {Vesperini}, E., {Sollima}, A., {et~al.} 2015, \bibinfo{title}{{Evolution of binary stars in multiple-population globular clusters},} \mnras, 449, 629, \dodoi{10.1093/mnras/stv306}

\bibitem[{M.~W. {Hosek} {et~al.}(2019){Hosek}, {Lu}, {Anderson}, {Najarro}, {Ghez}, {Morris}, {Clarkson}, \& {Albers}}]{Hosek2019}
{Hosek}, Jr., M.~W., {Lu}, J.~R., {Anderson}, J., {et~al.} 2019, \bibinfo{title}{{The Unusual Initial Mass Function of the Arches Cluster},} \apj, 870, 44, \dodoi{10.3847/1538-4357/aaef90}

\bibitem[{J.~D. {Hunter}(2007){Hunter}}]{Hunter2007matplotlib}
{Hunter}, J.~D. 2007, \bibinfo{title}{{Matplotlib: A 2D Graphics Environment},} Computing in Science and Engineering, 9, 90, \dodoi{10.1109/MCSE.2007.55}

\bibitem[{J.~R. {Hurley} {et~al.}(2007){Hurley}, {Aarseth}, \& {Shara}}]{Hurley+2007}
{Hurley}, J.~R., {Aarseth}, S.~J., \& {Shara}, M.~M. 2007, \bibinfo{title}{{The Core Binary Fractions of Star Clusters from Realistic Simulations},} \apj, 665, 707, \dodoi{10.1086/517879}

\bibitem[{P. {Hut} {et~al.}(1992{\natexlab{a}}){Hut}, {McMillan}, \& {Romani}}]{Hut+1992hardbinaries}
{Hut}, P., {McMillan}, S., \& {Romani}, R.~W. 1992{\natexlab{a}}, \bibinfo{title}{{The Evolution of a Primordial Binary Population in a Globular Cluster},} \apj, 389, 527, \dodoi{10.1086/171229}

\bibitem[{P. {Hut} {et~al.}(1992{\natexlab{b}}){Hut}, {McMillan}, {Goodman}, {Mateo}, {Phinney}, {Pryor}, {Richer}, {Verbunt}, \& {Weinberg}}]{Hut+1992review}
{Hut}, P., {McMillan}, S., {Goodman}, J., {et~al.} 1992{\natexlab{b}}, \bibinfo{title}{{Binaries in Globular Clusters},} \pasp, 104, 981, \dodoi{10.1086/133085}

\bibitem[{H.-C. {Hwang} {et~al.}(2021){Hwang}, {Ting}, {Schlaufman}, {Zakamska}, \& {Wyse}}]{Hwang+2021}
{Hwang}, H.-C., {Ting}, Y.-S., {Schlaufman}, K.~C., {Zakamska}, N.~L., \& {Wyse}, R. F.~G. 2021, \bibinfo{title}{{The non-monotonic, strong metallicity dependence of the wide-binary fraction},} \mnras, 501, 4329, \dodoi{10.1093/mnras/staa3854}

\bibitem[{A. {Hypki} {et~al.}(2025){Hypki}, {Vesperini}, {Giersz}, {Hong}, {Askar}, {Otulakowska-Hypka}, {Hellstrom}, \& {Wiktorowicz}}]{Hypki+2025}
{Hypki}, A., {Vesperini}, E., {Giersz}, M., {et~al.} 2025, \bibinfo{title}{{MOCCA: Global properties of tidally filling and underfilling globular star clusters with multiple stellar populations},} \aap, 693, A41, \dodoi{10.1051/0004-6361/202348653}

\bibitem[{J.~A. {Irwin} {et~al.}(2010){Irwin}, {Brink}, {Bregman}, \& {Roberts}}]{Irwin+2010}
{Irwin}, J.~A., {Brink}, T.~G., {Bregman}, J.~N., \& {Roberts}, T.~P. 2010, \bibinfo{title}{{Evidence for a Stellar Disruption by an Intermediate-mass Black Hole in an Extragalactic Globular Cluster},} \apjl, 712, L1, \dodoi{10.1088/2041-8205/712/1/L1}

\bibitem[{N. {Ivanova} {et~al.}(2005){Ivanova}, {Belczynski}, {Fregeau}, \& {Rasio}}]{Ivanova+2005}
{Ivanova}, N., {Belczynski}, K., {Fregeau}, J.~M., \& {Rasio}, F.~A. 2005, \bibinfo{title}{{The evolution of binary fractions in globular clusters},} \mnras, 358, 572, \dodoi{10.1111/j.1365-2966.2005.08804.x}

\bibitem[{N. {Ivanova} {et~al.}(2008){Ivanova}, {Heinke}, {Rasio}, {Belczynski}, \& {Fregeau}}]{Ivanova+2008}
{Ivanova}, N., {Heinke}, C.~O., {Rasio}, F.~A., {Belczynski}, K., \& {Fregeau}, J.~M. 2008, \bibinfo{title}{{Formation and evolution of compact binaries in globular clusters - II. Binaries with neutron stars},} \mnras, 386, 553, \dodoi{10.1111/j.1365-2966.2008.13064.x}

\bibitem[{T. {Je{\v{r}}{\'a}bkov{\'a}} {et~al.}(2017){Je{\v{r}}{\'a}bkov{\'a}}, {Kroupa}, {Dabringhausen}, {Hilker}, \& {Bekki}}]{Jerabkova2017}
{Je{\v{r}}{\'a}bkov{\'a}}, T., {Kroupa}, P., {Dabringhausen}, J., {Hilker}, M., \& {Bekki}, K. 2017, \bibinfo{title}{{The formation of ultra compact dwarf galaxies and massive globular clusters. Quasar-like objects to test for a variable stellar initial mass function},} \aap, 608, A53, \dodoi{10.1051/0004-6361/201731240}

\bibitem[{J. {Ji} \& J.~N. {Bregman}(2015){Ji} \& {Bregman}}]{Ji2015_gcbinaries}
{Ji}, J., \& {Bregman}, J.~N. 2015, \bibinfo{title}{{Binary Frequencies in a Sample of Globular Clusters. II. Sample Analysis and Comparison to Models},} \apj, 807, 32, \dodoi{10.1088/0004-637X/807/1/32}

\bibitem[{K.~J. {Joshi} {et~al.}(2000){Joshi}, {Rasio}, \& {Portegies Zwart}}]{Joshi2000}
{Joshi}, K.~J., {Rasio}, F.~A., \& {Portegies Zwart}, S. 2000, \bibinfo{title}{{Monte Carlo Simulations of Globular Cluster Evolution. I. Method and Test Calculations},} \apj, 540, 969, \dodoi{10.1086/309350}

\bibitem[{A.~W.~H. {Kamlah} {et~al.}(2022){Kamlah}, {Leveque}, {Spurzem}, {Arca Sedda}, {Askar}, {Banerjee}, {Berczik}, {Giersz}, {Hurley}, {Belloni}, {K{\"u}hmichel}, \& {Wang}}]{Kamlah+2022}
{Kamlah}, A.~W.~H., {Leveque}, A., {Spurzem}, R., {et~al.} 2022, \bibinfo{title}{{Preparing the next gravitational million-body simulations: evolution of single and binary stars in NBODY6++GPU, MOCCA, and MCLUSTER},} \mnras, 511, 4060, \dodoi{10.1093/mnras/stab3748}

\bibitem[{I. {King}(1962){King}}]{King1962cluster}
{King}, I. 1962, \bibinfo{title}{{The structure of star clusters. I. an empirical density law},} \aj, 67, 471, \dodoi{10.1086/108756}

\bibitem[{F. {K{\i}ro{\u{g}}lu} {et~al.}(2025){K{\i}ro{\u{g}}lu}, {Kremer}, \& {Rasio}}]{Kiroglu2025bhaccretion}
{K{\i}ro{\u{g}}lu}, F., {Kremer}, K., \& {Rasio}, F.~A. 2025, \bibinfo{title}{{Beyond Hierarchical Mergers: Accretion-driven Origins of Massive, Highly Spinning Black Holes in Dense Star Clusters},} \apjl, 994, L37, \dodoi{10.3847/2041-8213/ae1eeb}

\bibitem[{K. {Kremer}(2026){Kremer}}]{Kremer2026}
{Kremer}, K. 2026, in Encyclopedia of Astrophysics, Volume 3, Vol.~3, 458--472, \dodoi{10.1016/B978-0-443-21439-4.00103-6}

\bibitem[{K. {Kremer} {et~al.}(2018){Kremer}, {Chatterjee}, {Rodriguez}, \& {Rasio}}]{Kremer+2018}
{Kremer}, K., {Chatterjee}, S., {Rodriguez}, C.~L., \& {Rasio}, F.~A. 2018, \bibinfo{title}{{Accreting Black Hole Binaries in Globular Clusters},} \apj, 852, 29, \dodoi{10.3847/1538-4357/aa99df}

\bibitem[{K. {Kremer} {et~al.}(2019){Kremer}, {Chatterjee}, {Ye}, {Rodriguez}, \& {Rasio}}]{Kremer2019corecollapse}
{Kremer}, K., {Chatterjee}, S., {Ye}, C.~S., {Rodriguez}, C.~L., \& {Rasio}, F.~A. 2019, \bibinfo{title}{{How Initial Size Governs Core Collapse in Globular Clusters},} \apj, 871, 38, \dodoi{10.3847/1538-4357/aaf646}

\bibitem[{K. {Kremer} {et~al.}(2021){Kremer}, {Piro}, \& {Li}}]{Kremer+2021frbgc}
{Kremer}, K., {Piro}, A.~L., \& {Li}, D. 2021, \bibinfo{title}{{Dynamical Formation Channels for Fast Radio Bursts in Globular Clusters},} \apjl, 917, L11, \dodoi{10.3847/2041-8213/ac13a0}

\bibitem[{K. {Kremer} {et~al.}(2025){Kremer}, {Weatherford}, {Hopkins}, {Rui}, \& {Ye}}]{Kremer+2025}
{Kremer}, K., {Weatherford}, N.~C., {Hopkins}, P.~F., {Rui}, N.~Z., \& {Ye}, C.~S. 2025, \bibinfo{title}{{Connecting Cores and Black Hole Dynamics across Scales: From Globular Clusters to Massive Ellipticals},} \apjl, 993, L34, \dodoi{10.3847/2041-8213/ae1233}

\bibitem[{K. {Kremer} {et~al.}(2020){Kremer}, {Ye}, {Rui}, {Weatherford}, {Chatterjee}, {Fragione}, {Rodriguez}, {Spera}, \& {Rasio}}]{Kremer2020catalog}
{Kremer}, K., {Ye}, C.~S., {Rui}, N.~Z., {et~al.} 2020, \bibinfo{title}{{Modeling Dense Star Clusters in the Milky Way and Beyond with the CMC Cluster Catalog},} \apjs, 247, 48, \dodoi{10.3847/1538-4365/ab7919}

\bibitem[{P. {Kroupa}(1995){Kroupa}}]{Kroupa1995}
{Kroupa}, P. 1995, \bibinfo{title}{{The dynamical properties of stellar systems in the Galactic disc},} \mnras, 277, 1507, \dodoi{10.1093/mnras/277.4.1507}

\bibitem[{P. {Kroupa}(2001){Kroupa}}]{Kroupa2001imf}
{Kroupa}, P. 2001, \bibinfo{title}{{On the variation of the initial mass function},} \mnras, 322, 231, \dodoi{10.1046/j.1365-8711.2001.04022.x}

\bibitem[{M.~R. {Krumholz} {et~al.}(2019){Krumholz}, {McKee}, \& {Bland-Hawthorn}}]{Krumholz2019starclusters}
{Krumholz}, M.~R., {McKee}, C.~F., \& {Bland-Hawthorn}, J. 2019, \bibinfo{title}{{Star Clusters Across Cosmic Time},} \araa, 57, 227, \dodoi{10.1146/annurev-astro-091918-104430}

\bibitem[{N.~W.~C. {Leigh} {et~al.}(2015){Leigh}, {Giersz}, {Marks}, {Webb}, {Hypki}, {Heinke}, {Kroupa}, \& {Sills}}]{Leigh+2015}
{Leigh}, N. W.~C., {Giersz}, M., {Marks}, M., {et~al.} 2015, \bibinfo{title}{{The state of globular clusters at birth - II. Primordial binaries},} \mnras, 446, 226, \dodoi{10.1093/mnras/stu2110}

\bibitem[{N.~W.~C. {Leigh} {et~al.}(2013){Leigh}, {Giersz}, {Webb}, {Hypki}, {De Marchi}, {Kroupa}, \& {Sills}}]{Leigh+2013}
{Leigh}, N. W.~C., {Giersz}, M., {Webb}, J.~J., {et~al.} 2013, \bibinfo{title}{{The state of globular clusters at birth: emergence from the gas-embedded phase},} \mnras, 436, 3399, \dodoi{10.1093/mnras/stt1825}

\bibitem[{R. {Liu} {et~al.}(2025){Liu}, {Shao}, \& {Li}}]{Liu+2025}
{Liu}, R., {Shao}, Z., \& {Li}, L. 2025, \bibinfo{title}{{Mass-dependent Radial Distribution of Single and Binary Stars in the Pleiades and Their Dynamical Implications},} \apjl, 982, L43, \dodoi{10.3847/2041-8213/adbe60}

\bibitem[{J.~R. {Lu} {et~al.}(2013){Lu}, {Do}, {Ghez}, {Morris}, {Yelda}, \& {Matthews}}]{Lu+2013}
{Lu}, J.~R., {Do}, T., {Ghez}, A.~M., {et~al.} 2013, \bibinfo{title}{{Stellar Populations in the Central 0.5 pc of the Galaxy. II. The Initial Mass Function},} \apj, 764, 155, \dodoi{10.1088/0004-637X/764/2/155}

\bibitem[{T.~J. {Maccarone} {et~al.}(2007){Maccarone}, {Kundu}, {Zepf}, \& {Rhode}}]{Maccarone+2007}
{Maccarone}, T.~J., {Kundu}, A., {Zepf}, S.~E., \& {Rhode}, K.~L. 2007, \bibinfo{title}{{A black hole in a globular cluster},} \nat, 445, 183, \dodoi{10.1038/nature05434}

\bibitem[{A.~D. {Mackey} {et~al.}(2007){Mackey}, {Wilkinson}, {Davies}, \& {Gilmore}}]{Mackey+2007}
{Mackey}, A.~D., {Wilkinson}, M.~I., {Davies}, M.~B., \& {Gilmore}, G.~F. 2007, \bibinfo{title}{{The effect of stellar-mass black holes on the structural evolution of massive star clusters},} \mnras, 379, L40, \dodoi{10.1111/j.1745-3933.2007.00330.x}

\bibitem[{A.~D. {Mackey} {et~al.}(2008){Mackey}, {Wilkinson}, {Davies}, \& {Gilmore}}]{Mackey+2008}
{Mackey}, A.~D., {Wilkinson}, M.~I., {Davies}, M.~B., \& {Gilmore}, G.~F. 2008, \bibinfo{title}{{Black holes and core expansion in massive star clusters},} \mnras, 386, 65, \dodoi{10.1111/j.1365-2966.2008.13052.x}

\bibitem[{M.~A.~S. {Martinez} {et~al.}(2020){Martinez}, {Fragione}, {Kremer}, {Chatterjee}, {Rodriguez}, {Samsing}, {Ye}, {Weatherford}, {Zevin}, {Naoz}, \& {Rasio}}]{Martinez+2020}
{Martinez}, M. A.~S., {Fragione}, G., {Kremer}, K., {et~al.} 2020, \bibinfo{title}{{Black Hole Mergers from Hierarchical Triples in Dense Star Clusters},} \apj, 903, 67, \dodoi{10.3847/1538-4357/abba25}

\bibitem[{C.~D. {Matzner}(2024){Matzner}}]{Matzner2024}
{Matzner}, C.~D. 2024, \bibinfo{title}{{Intense Star Cluster Formation: Stellar Masses, the Mass Function, and the Fundamental Mass Scale},} \apjl, 975, L17, \dodoi{10.3847/2041-8213/ad85d4}

\bibitem[{S.~L.~W. {McMillan} {et~al.}(1987){McMillan}, {McDermott}, \& {Taam}}]{McMillan+1987}
{McMillan}, S. L.~W., {McDermott}, P.~N., \& {Taam}, R.~E. 1987, \bibinfo{title}{{The Formation and Evolution of Tidal Binary Stytems},} \apj, 318, 261, \dodoi{10.1086/165365}

\bibitem[{A.~P. {Milone} {et~al.}(2012){Milone}, {Piotto}, {Bedin}, {Aparicio}, {Anderson}, {Sarajedini}, {Marino}, {Moretti}, {Davies}, {Chaboyer}, {Dotter}, {Hempel}, {Mar{\'\i}n-Franch}, {Majewski}, {Paust}, {Reid}, {Rosenberg}, \& {Siegel}}]{Milone2012gcbinaryfraction}
{Milone}, A.~P., {Piotto}, G., {Bedin}, L.~R., {et~al.} 2012, \bibinfo{title}{{The ACS survey of Galactic globular clusters. XII. Photometric binaries along the main sequence},} \aap, 540, A16, \dodoi{10.1051/0004-6361/201016384}

\bibitem[{M. {Moe} \& R. {Di Stefano}(2017){Moe} \& {Di Stefano}}]{MoeDiStefano2017}
{Moe}, M., \& {Di Stefano}, R. 2017, \bibinfo{title}{{Mind Your Ps and Qs: The Interrelation between Period (P) and Mass-ratio (Q) Distributions of Binary Stars},} \apjs, 230, 15, \dodoi{10.3847/1538-4365/aa6fb6}

\bibitem[{M. {Moe} {et~al.}(2019){Moe}, {Kratter}, \& {Badenes}}]{Moe2019metallicity}
{Moe}, M., {Kratter}, K.~M., \& {Badenes}, C. 2019, \bibinfo{title}{{The Close Binary Fraction of Solar-type Stars Is Strongly Anticorrelated with Metallicity},} \apj, 875, 61, \dodoi{10.3847/1538-4357/ab0d88}

\bibitem[{M. {Morscher} {et~al.}(2015){Morscher}, {Pattabiraman}, {Rodriguez}, {Rasio}, \& {Umbreit}}]{Morscher2015bhdynamics}
{Morscher}, M., {Pattabiraman}, B., {Rodriguez}, C., {Rasio}, F.~A., \& {Umbreit}, S. 2015, \bibinfo{title}{{The Dynamical Evolution of Stellar Black Holes in Globular Clusters},} \apj, 800, 9, \dodoi{10.1088/0004-637X/800/1/9}

\bibitem[{M. {Morscher} {et~al.}(2013){Morscher}, {Umbreit}, {Farr}, \& {Rasio}}]{Morscher2013bhdynamics}
{Morscher}, M., {Umbreit}, S., {Farr}, W.~M., \& {Rasio}, F.~A. 2013, \bibinfo{title}{{Retention of Stellar-mass Black Holes in Globular Clusters},} \apjl, 763, L15, \dodoi{10.1088/2041-8205/763/1/L15}

\bibitem[{J. {M{\"u}ller-Horn} {et~al.}(2025){M{\"u}ller-Horn}, {G{\"o}ttgens}, {Dreizler}, {Kamann}, {Martens}, {Saracino}, \& {Ye}}]{MullerHorn2025_47tucbinaries}
{M{\"u}ller-Horn}, J., {G{\"o}ttgens}, F., {Dreizler}, S., {et~al.} 2025, \bibinfo{title}{{Binary properties of the globular cluster 47 Tuc (NGC 104): A dearth of short-period binaries},} \aap, 693, A161, \dodoi{10.1051/0004-6361/202450709}

\bibitem[{S.~S.~R. {Offner} {et~al.}(2023){Offner}, {Moe}, {Kratter}, {Sadavoy}, {Jensen}, \& {Tobin}}]{Offner2023binaries}
{Offner}, S.~S.~R., {Moe}, M., {Kratter}, K.~M., {et~al.} 2023, in Astronomical Society of the Pacific Conference Series, Vol. 534, Protostars and Planets VII, ed. S.~{Inutsuka}, Y.~{Aikawa}, T.~{Muto}, K.~{Tomida}, \& M.~{Tamura}, 275, \dodoi{10.48550/arXiv.2203.10066}

\bibitem[{L. {Paiella} {et~al.}(2025){Paiella}, {Ugolini}, {Spera}, {Branchesi}, \& {Arca Sedda}}]{Paiella+2025}
{Paiella}, L., {Ugolini}, C., {Spera}, M., {Branchesi}, M., \& {Arca Sedda}, M. 2025, \bibinfo{title}{{Assembling GW231123 in Star Clusters through the Combination of Stellar Binary Evolution and Hierarchical Mergers},} \apjl, 994, L54, \dodoi{10.3847/2041-8213/ae1447}

\bibitem[{B. {Pattabiraman} {et~al.}(2013){Pattabiraman}, {Umbreit}, {Liao}, {Choudhary}, {Kalogera}, {Memik}, \& {Rasio}}]{Pattabiraman2013}
{Pattabiraman}, B., {Umbreit}, S., {Liao}, W.-k., {et~al.} 2013, \bibinfo{title}{{A Parallel Monte Carlo Code for Simulating Collisional N-body Systems},} \apjs, 204, 15, \dodoi{10.1088/0067-0049/204/2/15}

\bibitem[{A. {Phillips} {et~al.}(2026){Phillips}, {Conroy}, {Nibauer}, {Wang}, {Chandra}, {Bonaca}, {Strader}, \& {MacLeod}}]{Phillips+2026}
{Phillips}, A., {Conroy}, C., {Nibauer}, J., {et~al.} 2026, \bibinfo{title}{{The Binary Populations of Stellar Streams are Set by Cluster Dynamics},} arXiv e-prints, arXiv:2603.06790, \dodoi{10.48550/arXiv.2603.06790}

\bibitem[{S.~F. {Portegies Zwart} {et~al.}(2010){Portegies Zwart}, {McMillan}, \& {Gieles}}]{ProtegiesZwart2010ymcreview}
{Portegies Zwart}, S.~F., {McMillan}, S. L.~W., \& {Gieles}, M. 2010, \bibinfo{title}{{Young Massive Star Clusters},} \araa, 48, 431, \dodoi{10.1146/annurev-astro-081309-130834}

\bibitem[{D. {Raghavan} {et~al.}(2010){Raghavan}, {McAlister}, {Henry}, {Latham}, {Marcy}, {Mason}, {Gies}, {White}, \& {ten Brummelaar}}]{Raghavan+2010}
{Raghavan}, D., {McAlister}, H.~A., {Henry}, T.~J., {et~al.} 2010, \bibinfo{title}{{A Survey of Stellar Families: Multiplicity of Solar-type Stars},} \apjs, 190, 1, \dodoi{10.1088/0067-0049/190/1/1}

\bibitem[{B.~W. {Ritchie} {et~al.}(2022){Ritchie}, {Clark}, {Negueruela}, \& {Najarro}}]{Ritchie2022w1obbinaries}
{Ritchie}, B.~W., {Clark}, J.~S., {Negueruela}, I., \& {Najarro}, F. 2022, \bibinfo{title}{{A VLT/FLAMES survey for massive binaries in Westerlund 1. VIII. Binary systems and orbital parameters},} \aap, 660, A89, \dodoi{10.1051/0004-6361/202142405}

\bibitem[{C.~L. {Rodriguez} {et~al.}(2015){Rodriguez}, {Morscher}, {Pattabiraman}, {Chatterjee}, {Haster}, \& {Rasio}}]{Rodriguez2015}
{Rodriguez}, C.~L., {Morscher}, M., {Pattabiraman}, B., {et~al.} 2015, \bibinfo{title}{{Binary Black Hole Mergers from Globular Clusters: Implications for Advanced LIGO},} \prl, 115, 051101, \dodoi{10.1103/PhysRevLett.115.051101}

\bibitem[{C.~L. {Rodriguez} {et~al.}(2022){Rodriguez}, {Weatherford}, {Coughlin}, {Amaro-Seoane}, {Breivik}, {Chatterjee}, {Fragione}, {K{\i}ro{\u{g}}lu}, {Kremer}, {Rui}, {Ye}, {Zevin}, \& {Rasio}}]{Rodriguez2022cmcreview}
{Rodriguez}, C.~L., {Weatherford}, N.~C., {Coughlin}, S.~C., {et~al.} 2022, \bibinfo{title}{{Modeling Dense Star Clusters in the Milky Way and beyond with the Cluster Monte Carlo Code},} \apjs, 258, 22, \dodoi{10.3847/1538-4365/ac2edf}

\bibitem[{H. {Sana} {et~al.}(2012){Sana}, {de Mink}, {de Koter}, {Langer}, {Evans}, {Gieles}, {Gosset}, {Izzard}, {Le Bouquin}, \& {Schneider}}]{Sana+2012}
{Sana}, H., {de Mink}, S.~E., {de Koter}, A., {et~al.} 2012, \bibinfo{title}{{Binary Interaction Dominates the Evolution of Massive Stars},} Science, 337, 444, \dodoi{10.1126/science.1223344}

\bibitem[{H. {Sana} {et~al.}(2013){Sana}, {de Koter}, {de Mink}, {Dunstall}, {Evans}, {H{\'e}nault-Brunet}, {Ma{\'\i}z Apell{\'a}niz}, {Ram{\'\i}rez-Agudelo}, {Taylor}, {Walborn}, {Clark}, {Crowther}, {Herrero}, {Gieles}, {Langer}, {Lennon}, \& {Vink}}]{Sana2013_30Dor}
{Sana}, H., {de Koter}, A., {de Mink}, S.~E., {et~al.} 2013, \bibinfo{title}{{The VLT-FLAMES Tarantula Survey. VIII. Multiplicity properties of the O-type star population},} \aap, 550, A107, \dodoi{10.1051/0004-6361/201219621}

\bibitem[{H. {Sana} {et~al.}(2025){Sana}, {Shenar}, {Bodensteiner}, {Britavskiy}, {Langer}, {Lennon}, {Mahy}, {Mandel}, {de Mink}, {Patrick}, {Villase{\~n}or}, {Dirickx}, {Abdul-Masih}, {Almeida}, {Backs}, {Berlanas}, {Bernini-Peron}, {Bowman}, {Bronner}, {Crowther}, {Deshmukh}, {Evans}, {Fabry}, {Gieles}, {Gilkis}, {Gonz{\'a}lez-Tor{\`a}}, {Gr{\"a}fener}, {G{\"o}tberg}, {Hawcroft}, {H{\'e}nault-Brunet}, {Herrero}, {Holgado}, {Izzard}, {de Koter}, {Janssens}, {Johnston}, {Josiek}, {Justham}, {Kalari}, {Klencki}, {Kub{\'a}t}, {Kub{\'a}tov{\'a}}, {Lefever}, {van Loon}, {Ludwig}, {Mackey}, {Ma{\'\i}z Apell{\'a}niz}, {Maravelias}, {Marchant}, {Mazeh}, {Menon}, {Moe}, {Najarro}, {Oskinova}, {Ovadia}, {Pauli}, {Pawlak}, {Ramachandran}, {Renzo}, {Rocha}, {Sander}, {Schneider}, {Schootemeijer}, {Sch{\"o}sser}, {Sch{\"u}rmann}, {Sen}, {Shahaf}, {Sim{\'o}n-D{\'\i}az}, {van Son}, {Stoop}, {Toonen}, {Tramper}, {Valli}, {Vigna-G{\'o}mez}, {Vink}, {Wang}, \& {Willcox}}]{Sana2025bloem}
{Sana}, H., {Shenar}, T., {Bodensteiner}, J., {et~al.} 2025, \bibinfo{title}{{A high fraction of close massive binary stars at low metallicity},} Nature Astronomy, 9, 1337, \dodoi{10.1038/s41550-025-02610-x}

\bibitem[{R.~A. {Scheepmaker} {et~al.}(2007){Scheepmaker}, {Haas}, {Gieles}, {Bastian}, {Larsen}, \& {Lamers}}]{Scheepmaker+2007}
{Scheepmaker}, R.~A., {Haas}, M.~R., {Gieles}, M., {et~al.} 2007, \bibinfo{title}{{ACS imaging of star clusters in M 51. I. Identification and radius distribution},} \aap, 469, 925, \dodoi{10.1051/0004-6361:20077511}

\bibitem[{A. {Sollima} {et~al.}(2007){Sollima}, {Beccari}, {Ferraro}, {Fusi Pecci}, \& {Sarajedini}}]{Sollima+2007}
{Sollima}, A., {Beccari}, G., {Ferraro}, F.~R., {Fusi Pecci}, F., \& {Sarajedini}, A. 2007, \bibinfo{title}{{The fraction of binary systems in the core of 13 low-density Galactic globular clusters},} \mnras, 380, 781, \dodoi{10.1111/j.1365-2966.2007.12116.x}

\bibitem[{R. {Spurzem} \& A. {Kamlah}(2023){Spurzem} \& {Kamlah}}]{SpurzemKamlah2023}
{Spurzem}, R., \& {Kamlah}, A. 2023, \bibinfo{title}{{Computational methods for collisional stellar systems},} Living Reviews in Computational Astrophysics, 9, 3, \dodoi{10.1007/s41115-023-00018-w}

\bibitem[{J. {Strader} {et~al.}(2012){Strader}, {Chomiuk}, {Maccarone}, {Miller-Jones}, \& {Seth}}]{Strader+2012}
{Strader}, J., {Chomiuk}, L., {Maccarone}, T.~J., {Miller-Jones}, J. C.~A., \& {Seth}, A.~C. 2012, \bibinfo{title}{{Two stellar-mass black holes in the globular cluster M22},} \nat, 490, 71, \dodoi{10.1038/nature11490}

\bibitem[{ {The LIGO Scientific Collaboration} {et~al.}(2026){The LIGO Scientific Collaboration}, {the Virgo Collaboration}, {the KAGRA Collaboration}, {Abac}, {Abouelfettouh}, {Acernese}, {Ackley}, {Adamcewicz}, {Adhicary}, {Adhikari}, {Adhikari}, {Adhikari}, {Adkins}, {Afroz}, {Agapito}, {Agarwal}, {Agathos}, {Aggarwal}, {Aggarwal}, {Aguiar}, {Ahrend}, {Aiello}, {Ain}, {Ajith}, {Akutsu}, {Albanesi}, {Ali}, {Al-Kershi}, {All{\'e}n{\'e}}, {Allocca}, {Al-Shammari}, {Altin}, {Alvarez-Lopez}, {Amar}, {Amarasinghe}, {Amato}, {Amicucci}, {Amra}, {Ananyeva}, {Anderson}, {Anderson}, {Andia}, {Ando}, {Andr{\'e}s-Carcasona}, {Andri{\'c}}, {Anglin}, {Ansoldi}, {Antelis}, {Antier}, {Aoumi}, {Appavuravther}, {Appert}, {Apple}, {Arai}, {Araya}, {Araya}, {Arca Sedda}, {Areeda}, {Aritomi}, {Armato}, {Armstrong}, {Arnaud}, {Arogeti}, {Aronson}, {Arun}, {Ashton}, {Aso}, {Asprea}, {Assiduo}, {Assis de Souza Melo}, {Aston}, {Astone}, {Attadio}, {Aubin}, {AultONeal}, {Avallone}, {Avila}, {Babak}, {Badger}, {Bae}, {Bagnasco},
  {Baiotti}, {Bajpai}, {Baka}, {Baker}, {Baker}, {Baker}, {Baldi}, {Baldicchi}, {Ball}, {Ballardin}, {Ballmer}, {Banagiri}, {Banerjee}, {Bankar}, {Baptiste}, {Baral}, {Baratti}, {Barayoga}, {Barish}, {Barker}, {Barman}, {Barneo}, {Barone}, {Barr}, {Barsotti}, {Barsuglia}, {Barta}, {Bartoletti}, {Barton}, {Bartos}, {Basalaev}, {Bassiri}, {Basti}, {Bawaj}, {Baxi}, {Bayley}, {Baylor}, {Baynard}, {Bazzan}, {Bedakihale}, {Beirnaert}, {Bejger}, {Belardinelli}, {Bell}, {Bellie}, {Bellizzi}, {Benoit}, {Bentara}, {Bentley}, {Ben Yaala}, {Bera}, {Bergamin}, {Berger}, {Bernuzzi}, {Beroiz}, {Berry}, {Bersanetti}, {Bertheas}, {Bertolini}, {Betzwieser}, {Beveridge}, {Bevilacqua}, {Bevins}, {Bhagwat}, {Bhandare}, {Bhat}, {Bhatt}, {Bhattacharjee}, {Bhattacharyya}, {Bhaumik}, {Biancalana}, {Bianchi}, {Bilenko}, {Billingsley}, {Binetti}, {Bini}, {Binu}, {Biot}, {Birnholtz}, {Biscoveanu}, {Bisht}, {Bitossi}, {Bizouard}, {Blaber}, {Blackburn}, {Blagg}, {Blair}, {Blair}, {Bode}, {Boettner}, {Boileau}, {Boldrini}, {Bolingbroke},
  {Bolliand}, {Bonavena}, {Bondarescu}, {Bondu}, {Bonilla}, {Bonilla}, {Bonino}, {Bonnand}, {Borchers}, {Boschi}, {Bose}, {Bossilkov}, {Bothra}, {Boudon}, {Bourg}, {Boyle}, {Bozzi}, {Bradaschia}, {Brady}, {Branch}, {Branchesi}, {Braun}, {Briant}, {Brillet}, {Brinkmann}, \& {Brockill}}]{LVK2026_GWTC4}
{The LIGO Scientific Collaboration}, {the Virgo Collaboration}, {the KAGRA Collaboration}, {et~al.} 2026, \bibinfo{title}{{GWTC-4.0: Tests of General Relativity. I. Overview and General Tests},} arXiv e-prints, arXiv:2603.19019, \dodoi{10.48550/arXiv.2603.19019}

\bibitem[{C.~A. {Tout} {et~al.}(1996){Tout}, {Pols}, {Eggleton}, \& {Han}}]{Tout+1996}
{Tout}, C.~A., {Pols}, O.~R., {Eggleton}, P.~P., \& {Han}, Z. 1996, \bibinfo{title}{{Zero-age main-seqence radii and luminosities as analytic functions of mass and metallicity},} \mnras, 281, 257, \dodoi{10.1093/mnras/281.1.257}

\bibitem[{S.~A. {Usman} {et~al.}(2024){Usman}, {Ji}, {Li}, {Pace}, {Cullinane}, {Da Costa}, {Koposov}, {Lewis}, {Zucker}, {Belokurov}, {Bland-Hawthorn}, {Ferguson}, {Hansen}, {Limberg}, {Martell}, {McKenzie}, {Simon}, \& {S5 Collaboration}}]{UsmanJi2024gcstreams}
{Usman}, S.~A., {Ji}, A.~P., {Li}, T.~S., {et~al.} 2024, \bibinfo{title}{{Multiple populations and a CH star found in the 300S globular cluster stellar stream},} \mnras, 529, 2413, \dodoi{10.1093/mnras/stae185}

\bibitem[{P. {van Dokkum} \& C. {Conroy}(2024){van Dokkum} \& {Conroy}}]{vanDokkumConroy2024}
{van Dokkum}, P., \& {Conroy}, C. 2024, \bibinfo{title}{{Reconciling M/L Ratios Across Cosmic Time: a Concordance IMF for Massive Galaxies},} \apjl, 973, L32, \dodoi{10.3847/2041-8213/ad77b8}

\bibitem[{L. {Wang} {et~al.}(2020){Wang}, {Iwasawa}, {Nitadori}, \& {Makino}}]{Wang2020petar}
{Wang}, L., {Iwasawa}, M., {Nitadori}, K., \& {Makino}, J. 2020, \bibinfo{title}{{PETAR: a high-performance N-body code for modelling massive collisional stellar systems},} \mnras, 497, 536, \dodoi{10.1093/mnras/staa1915}

\bibitem[{L. {Wang} {et~al.}(2022){Wang}, {Tanikawa}, \& {Fujii}}]{Wang+2022}
{Wang}, L., {Tanikawa}, A., \& {Fujii}, M.~S. 2022, \bibinfo{title}{{The impact of primordial binary on the dynamical evolution of intermediate massive star clusters},} \mnras, 509, 4713, \dodoi{10.1093/mnras/stab3255}

\bibitem[{N.~C. {Weatherford} \& A. {Bonaca}(2026){Weatherford} \& {Bonaca}}]{Weatherford+2026}
{Weatherford}, N.~C., \& {Bonaca}, A. 2026, \bibinfo{title}{{Kinematics of Stellar Streams from Globular Clusters Depend on Black Hole Retention and Star Mass: A Selection Effect for Dark Matter Inference},} \apj, 997, 90, \dodoi{10.3847/1538-4357/ae21e0}

\bibitem[{N.~C. {Weatherford} {et~al.}(2020){Weatherford}, {Chatterjee}, {Kremer}, \& {Rasio}}]{Weatherford+2020}
{Weatherford}, N.~C., {Chatterjee}, S., {Kremer}, K., \& {Rasio}, F.~A. 2020, \bibinfo{title}{{A Dynamical Survey of Stellar-mass Black Holes in 50 Milky Way Globular Clusters},} \apj, 898, 162, \dodoi{10.3847/1538-4357/ab9f98}

\bibitem[{N.~C. {Weatherford} {et~al.}(2023){Weatherford}, {K{\i}ro{\u{g}}lu}, {Fragione}, {Chatterjee}, {Kremer}, \& {Rasio}}]{Weatherford+2023}
{Weatherford}, N.~C., {K{\i}ro{\u{g}}lu}, F., {Fragione}, G., {et~al.} 2023, \bibinfo{title}{{Stellar Escape from Globular Clusters. I. Escape Mechanisms and Properties at Ejection},} \apj, 946, 104, \dodoi{10.3847/1538-4357/acbcc1}

\bibitem[{K. {Wu} {et~al.}(2025){Wu}, {Cho}, {Spurzem}, {Wang}, {Flammini Dotti}, \& {Amiri}}]{Wu+2025}
{Wu}, K., {Cho}, P., {Spurzem}, R., {et~al.} 2025, \bibinfo{title}{{DRAGON-III simulation: modelling million-body globular and nuclear star clusters},} arXiv e-prints, arXiv:2510.03933, \dodoi{10.48550/arXiv.2510.03933}

\bibitem[{W. {Wu} {et~al.}(2026){Wu}, {Kroupa}, \& {Jadhav}}]{Wu+2026petarocbinaries}
{Wu}, W., {Kroupa}, P., \& {Jadhav}, V.~V. 2026, \bibinfo{title}{{Statistical study for binary star evolution in dense embedded clusters},} arXiv e-prints, arXiv:2601.22767, \dodoi{10.48550/arXiv.2601.22767}

\bibitem[{C.~S. {Ye} {et~al.}(2019){Ye}, {Kremer}, {Chatterjee}, {Rodriguez}, \& {Rasio}}]{Ye+2019mspgc}
{Ye}, C.~S., {Kremer}, K., {Chatterjee}, S., {Rodriguez}, C.~L., \& {Rasio}, F.~A. 2019, \bibinfo{title}{{Millisecond Pulsars and Black Holes in Globular Clusters},} \apj, 877, 122, \dodoi{10.3847/1538-4357/ab1b21}

\bibitem[{C.~S. {Ye} {et~al.}(2022){Ye}, {Kremer}, {Rodriguez}, {Rui}, {Weatherford}, {Chatterjee}, {Fragione}, \& {Rasio}}]{Ye+2022}
{Ye}, C.~S., {Kremer}, K., {Rodriguez}, C.~L., {et~al.} 2022, \bibinfo{title}{{Compact Object Modeling in the Globular Cluster 47 Tucanae},} \apj, 931, 84, \dodoi{10.3847/1538-4357/ac5b0b}

\end{thebibliography}
\bibliographystyle{aasjournalv7}

\end{document}